\documentclass[aps,groupedaddress,showpacs,letterpaper,twocolumn]{revtex4}
\usepackage{graphicx} 
\usepackage{amsmath} 
\usepackage{amsfonts}
\usepackage{array}
\newcommand{\Rb}{ \mbox{$^{87}$}Rb }

\newcommand \bfalpha{{\boldsymbol \alpha}}
\newcommand \bfepsilon{{\boldsymbol \epsilon}}
\newcommand \mcE{{\mathcal E}}

\begin{document}
\title{Analysis of a quantum logic device based on dipole-dipole interactions of optically trapped Rydberg atoms }
\author{M. Saffman and T. G. Walker }
\affiliation{
Department of Physics,
University of Wisconsin, 1150 University Avenue,  Madison, Wisconsin 53706
}
 \date{\today}

\begin{abstract}
We present a detailed analysis and design of a neutral atom quantum logic device based on atoms in optical 
traps interacting via dipole-dipole
coupling of Rydberg states. The dominant physical mechanisms leading to 
decoherence and loss of fidelity are  enumerated. 
Our results support the feasibility of performing  single and two-qubit gates at MHz rates with 
decoherence probability and fidelity errors at the level of  $10^{-3}$ for each operation. 
Current limitations and
possible approaches to further improvement of
the device are discussed.  
\end{abstract}

\maketitle

\section{Introduction}

Motivated by the discovery that quantum algorithms can  provide exponential gains for solving certain computational problems, numerous proposals have been advanced for experimental realization of a quantum computer\cite{ref.qcbook}. While a useful processor remains far off, ground breaking experiments have demonstrated controlled evolution of a few qubits and implemented basic quantum algorithms for computation and error correction\cite{ref.kimbleqed,ref.monroe,ref.sackett,ref.chuang,ref.kwiat,ref.knill,ref.winelanderror}.  Among the range of  physical systems 
that have been identified as candidates for implementing quantum logic 
the most extensive laboratory results so far have been obtained with cold trapped ions\cite{winelandandblatt} and nuclear magnetic resonance (NMR) in macroscopic 
samples\cite{ref.chuang2,ref.cory}.

Within the last few years neutral atoms have emerged as a possible route to experimental quantum logic. 
The most obvious distinguishing feature between neutral atom and trapped ion schemes is the absence of strong Coulomb forces in the former. Coulomb forces between ions couple strongly the motional degrees of freedom.  This can be utilized to entangle
any two of a linear string of ions, as was first elucidated in the work of Cirac and Zoller\cite{ref.ciraczoller} and demonstrated experimentally 
in Boulder\cite{ref.wineland} and Innsbruck\cite{ref.blatt}.   The lack of a strong Coulomb interaction in neutral atoms is advantageous as regards decoherence, since  coupling to stray fields is weaker for atoms than for ions.
The drawback, and indeed the central difficulty in constructing a large scale quantum processor, is the need for 
strong qubit to qubit coupling, while maintaining weak coupling to the environment. 
Neutral atom coupling based on ground state collisions, light mediated dipole-dipole coupling, and dipole-dipole coupling of highly excited Rydberg states have all been proposed in the last several years\cite{ref.brennen,ref.jaksch,ref.deutsch,ref.calarco,ref.qcatomsionslight,ref.motionalgate}. In particular dipole-dipole coupling of Rydberg states provides a strong interaction suitable for the implementation of fast gates\cite{ref.cote}, and this paper is devoted to a detailed study of this approach.

While the theoretical foundations of the Jaksch et al. Rydberg state dipole-dipole coupling approach to quantum logic have been presented\cite{ref.cote} the question of how to implement this scheme in a practical and scalable fashion has not been solved\cite{ref.grangier}. Regardless of how logical operations are to be performed, there are two primary 
obstacles that must be surmounted. The first is how to create a large number of trapping sites, and load a single atom into each site. This amounts to initialization of the quantum computer. The second difficulty is that in order to be useful for generic models of quantum computation the sites must be individually accessible for logical operations, and state readout.  In this paper we do not discuss the problem of creation and loading of a large number of addressable single atom sites. A number of possible solutions to these questions   
have been discussed in the literature\cite{ref.mott,ref.porto,ref.weiss,ref.zoller,ref.hanschco2lattice,ref.ertmer,ref.sw,ref.besselimaging}. 

Our goal here is to examine in detail the use of two closely spaced sites, each containing a neutral atom qubit, for high fidelity quantum operations. 
Far-off-resonance optical traps (FORTs) are  defined by tightly focusing laser light in a set of chosen  locations. Single atoms are loaded into the optical traps after precooling in a 
magneto-optical trap\cite{ref.meschedesingleatom,ref.grangiernature}. Single qubit operations are performed using two-photon stimulated Raman transitions, and a two-qubit conditional phase gate is realized using dipole-dipole coupling of atoms excited to high lying Rydberg states\cite{ref.cote}.
Qubit measurement is performed by counting resonance fluorescence photons.

The ability to perform many operations with
high fidelity and low decoherence is a prerequisite for scaling up to a larger number of qubits. As will be shown in what follows our calculations lead to the conclusion that a set of logically complete qubit operations can be performed with high fidelity at MHz rates. This would suggest that qubit storage in  optical traps with coherence times of much less than a second will be sufficient for large  computations. However, the necessity of implementing error correction implies that a computation will also require a large number of state measurements, which are projected to be several orders of magnitude slower than the logical operations. We therefore examine closely the feasibility 
of $T1$ and $T2$ coherence times of several seconds in optical traps.

The remainder of this paper is structured as follows. In Sec. \ref{sec.singleatomtraps} we estimate the decoherence times for storage of individual atoms in FORTs. We specifically include the contributions due to
 collisions with hot background atoms (Sec. \ref{sec.trapbackground}), 
photon scattering from the trapping laser (Sec. \ref{sec.trapscattering}), 
spin flips due to the trapping laser, and heating rates due to laser noise (Sec. \ref{sec.traplasernoise}). Decoherence due to fluctuations in the trapping lasers is considered in Sec. \ref{sec.acstarkshift} and due to background 
fields in Sec. \ref{sec.backgroundfields}.

In Sec. \ref{sec.qubit1} we discuss the operation of single qubit gates based on two-photon stimulated Raman transitions. In particular we calculate decoherence probabilities due to excited state spontaneous emission (Sec. \ref{sec.qubit1spe}) and expected gate fidelities due to 
ac Stark shifts(Sec. \ref{sec.acstarkshifts}) and motional effects   (Sec. \ref{sec.atomiclocalization}).
Leakage out of the computational basis due to imperfect optical polarization is estimated in Sec. \ref{sec.qubitpolarization} and limitations imposed by the laser stability are estimated in 
Sec. \ref{sec.qubit1lasernoise}.  The ability to rapidly interrogate the atomic state is crucial to the usefulness of this approach. We discuss single atom state detection using collection of resonance fluorescence in Sec. \ref{sec.qubit1fluorescence}. Included in Sec. \ref{sec.qubit1fluorescence} is a discussion of heating during readout, and its amelioration using red-detuned molasses for the interrogation beams. 

In Sec. \ref{sec.qubit2} we discuss the implementation of a two-qubit conditional phase gate which can serve as a logical primitive for arbitrary computations. We consider two different regimes of operation:
Rabi frequency large compared to dipole-dipole frequency shift (Sec. \ref{sec.qubit2rabilarge}) and dipole-dipole frequency shift large compared to Rabi frequency (Sec. \ref{sec.qubit2ddlarge}). 

Two-qubit operations involving Rydberg states have larger errors and higher decoherence rates than single qubit operations. 
We optimize the parameters of a phase gate in the two limits of weak dipole-dipole interaction  (Sec. \ref{sec.qubit2rabilarge}) and strong dipole-dipole 
interaction (Sec. \ref{sec.qubit2ddlarge}). In both cases the performance depends critically on the Rydberg state lifetime which we calculate for relevant experimental parameters in Secs. \ref{sec.qubit2radiative},\ref{sec.qubit2phi}. An additional aspect of the Rydberg state interactions that needs to be addressed is the rate of heating due to differences in the ground state and excited state polarizabilities. We show how to minimize this effect at the expense of some additional decoherence in Sec. (\ref{sec.balancepolarize}).

The results of the calculations of fidelities and decoherence rates provide a picture of the feasibility of a quantum logic device capable of executing a large number of sequential gate operations. We discuss the expected overall performance of this approach to quantum computing in Sec. 
\ref{sec.discussion}, and highlight the areas that are most troublesome. Possible extensions to the techniques  discussed here that have the potential for improved performance 
 are discussed.

\section{Optical traps for single atoms}
\label{sec.singleatomtraps}

In this section we recall some basic features of far off resonant optical traps. A number of distinct physical mechanisms limit the coherence of atoms stored in FORT's. Some of these decohering effects are intrinsic to the operation of the FORT, and some are due to technical imperfections of the apparatus used. As shown in Table \ref{tab.groundstate}  these mechanisms contribute to an effective decoherence time of diagonal ($T1$)  or off-diagonal  ($T2$) density matrix elements. 
We discuss the physics behind each of these decohering mechanisms in the following subsections. All numerical estimates of decoherence rates will be calculated for $^{87}$Rb atoms using the parameters listed in Table \ref{tab.groundstate}.

\begin{table}[t]
\centering
\begin{tabular}{l|c||c|c}
\hline
mechanism & Section & $T1 [\rm sec]$ & $T2 [\rm sec]$\\
\hline
Background gas collisions & \ref{sec.trapbackground}  & 55  & \\
Rayleigh scattering &\ref{sec.trapscattering}  & 97&\\
Raman scattering &\ref{sec.trapscattering}  &151&\\
Laser noise heating &\ref{sec.traplasernoise}  & 20&\\
AC stark shifts - intensity noise &\ref{sec.acstarkshift} &&12\\
AC stark shifts - motional &\ref{sec.acstarkshift} &&2.6\\
Background $B$ field & \ref{sec.backgroundfields}& &56\\
\hline
Combined & & 12 & 2.1\\
\hline
\end{tabular}
\caption{Physical mechanisms limiting groundstate coherence of qubit basis states $|a\rangle=|F=1,m_F=0\rangle$ and $|b\rangle=|F=2,m_F=0\rangle.$ Values listed are for $P= 1 \times 10^{-10}$ mbar, $U_m= 1$ mK, $T_a=50~\mu\rm K,$ $\lambda_f=1.01~\mu\rm m,$ and $w_{f0}=2.5~\mu\rm m$, and $\alpha_0(\omega_f)=114$
\AA$^3$.  See text for details and additional parameters. }
\label{tab.groundstate}
\end{table}

\subsection{FORT trap parameters}

In its simplest form an optical FORT trap can be created by focusing a single laser beam of wavelength $\lambda_f$ to a waist $w_{f0}$\cite{ref.chu86,ref.heinzen93}. 
The ground state AC Stark shift due to a far-detuned trapping beam is 
\begin{equation}
U_{\rm ac}=-\frac{1}{4}|\mcE_f|^2 
\bfepsilon^*\cdot\langle i|\hat\bfalpha| i\rangle\cdot\bfepsilon
\label{eq.acstark1}
\end{equation} 
where ${\mathcal E}_f$ is the amplitude of the optical field, the laser intensity is 
$I_f=\epsilon_0 c|{\mathcal E}_f|^2/2$, and $\bfepsilon$ is the trapping laser polarization. 
The polarizability is in general the sum of scalar, vector, and tensor parts\cite{ref.boninbook}. For a $J=1/2$ ground state with a linearly polarized trapping beam we need only consider the scalar polarizability which we calculate numerically using Coulomb wave functions to be 
$\alpha_0=114~ \AA^3$ for a trapping laser at 
$\lambda_f=1.01~\mu\rm m.$

The maximum depth of the potential well at the center of the focused beam expressed in temperature units is $U_m=-\alpha_0|{\mathcal E}_f|^2/4.$    
The spatial dependence of the trapping potential is then
\begin{equation}
U_f(x,y,z) = U_m \frac{e^{-2(x^2+y^2)/w_f^2(z) }}{1+\frac{z^2}{L_f^2}},
\label{eq.fort1}
\end{equation}
where for a  FORT beam propagating along $z,$ $w_f^2(z)=w_{f0}^2(1+z^2/L_f^2)$, with  the Rayleigh length $L_f=\pi w_{f0}^2/\lambda_f.$ For the parameters of Table \ref{tab.groundstate} a laser power of $60~\rm mW$ gives $U_m=1~\rm mK.$

The FORT can be directly loaded from a MOT provided that the product of the capture volume and the MOT density is larger than unity. 
A first approximation for the capture volume assumes that all atoms in the region where $|U_f/k_B|>T_c$ are captured, while those outside this region are lost (in the rest of the paper we will express all energies in temperature units and put $k_B=1$). We expect that the capture temperature $T_c$ will be similar to $T_a,$ the kinetic temperature of the atoms in the MOT, provided $T_a$ is smaller than $|U_m|.$ We define a relative trap depth $\chi=|U_m|/T_a,$ so that the capture volume vanishes for $\chi=1$, and  increases  
monotonically with increasing $\chi.$ A simple calculation then results in an expression for the capture volume,
\begin{eqnarray}
V&=& 2 \int_0^{z_m} dz ~ \pi r_m^2(z)
\nonumber\\
&=&\pi w_{f0}^2\int_0^{z_m} dz \left(1+\frac{z^2}{L_f^2} \right)\ln\left(\frac{z_m^2+L_f^2}{z^2+L_f^2} \right) \nonumber \\
&=& \frac{4\pi}{3}w_{f0}^2 z_m + 
\frac{2\pi}{9}\frac{w_{f0}^2}{L_f^2}\left[z_m^3 -6 L_f^3\tan^{-1}\left(\frac{z_m}{L_f} \right) \right]
\label{eq.fortvolume}
\end{eqnarray}
with $z_m=L_f\sqrt{\chi-1}.$
Using  numerical values from Table \ref{tab.groundstate}
 we get a capture volume of $8.6\times 10^{-9}~\rm cm^3.$ We have found in unpublished experiments that a  MOT density of a few times $10^8/\rm cm^3$ is sufficient to load single atoms as was  demonstrated by several groups
in recent years\cite{ref.grangiernature,ref.meschedesingleatom}.

In the context of quantum logic it is important that the atoms are well localized in position and momentum. We can estimate the variances of the atomic position and momentum by making a parabolic expansion of the FORT potential about the origin. The effective spring constants of the trap are found to be
\begin{subequations}
\begin{eqnarray}
\kappa_x=\kappa_y&=& 4 \frac{|U_m|}{w_{f0}^2}\\
\kappa_z&=& 2 \frac{ |U_m|}{L_f^2}
\end{eqnarray}
  \end{subequations}
and the corresponding oscillation frequencies are 
\begin{subequations}
\begin{eqnarray}
\omega_x=\omega_y&=&\frac{2}{w_{f0} } \left(\frac{|U_m|}{m}\right)^{1/2}\\
\omega_z&=&\frac{\sqrt 2}{L_f} \left(\frac{|U_m|}{m}\right)^{1/2},
\end{eqnarray}
\label{eq.trapvibration}
\end{subequations}
\noindent
with $m$ the atomic mass. For the above parameters and $^{87}$Rb 
we find $\omega_{(x,z)}/(2\pi) =(39,3.6) ~\rm kHz$. At $T_a=50~\mu\rm K$ many vibrational levels will be excited and we can use Boltzmann factors to estimate the time averaged variances of the position and momentum as 
\begin{subequations}
\begin{eqnarray}
\left<x_a^2 \right>=\left<y_a^2 \right>&=& 
\frac{w_{f0}^2}{4}\frac{T_a}{|U_m|},
\label{eq.xsq}\\
\left<z_a^2 \right>&=& 
\frac{\pi^2w_{f0}^4}{2\lambda_f^2}\frac{T_a}{|U_m|},\\
\left< v_{xa}^2\right>=\left< v_{ya}^2\right>=\left< v_{za}^2\right>&=&\frac{T_a}{m}.
\label{eq.trapmotion}
\end{eqnarray}
  \end{subequations}
Note that the spatial localization along $z$ can be written in terms of an anisotropy factor $\xi_f=\sqrt{\kappa_x/\kappa_z}=\pi w_{f0}/\lambda_f$ such 
that $\left<z_a^2 \right>=\xi_f^2 \left<x_a^2 \right>.$

\subsection{Background gas collisions}
\label{sec.trapbackground}

Collisions with untrapped background atoms in the vacuum chamber result in heating and loss of atoms from the FORT and therefore limit the storage time and T1 that can be achieved. The characteristic  energy change for which diffractive collisions must be accounted for is  $\sim2.8~\rm mK$ for Rb\cite{ref.bali99}. As we are considering a  shallow FORT of  depth  $|U_m|=1~\rm mK$ we will neglect diffractive and heating 
effects and approximate the FORT lifetime due to background collisions mediated by a van der Waals interaction as\cite{ref.bali99}  
\begin{equation}
\frac{1}{\tau_c} = \sqrt{\frac{3  T_b}{m}} n_b\sigma_{\rm Rb-Rb} 
\end{equation}
with $T_b=300~\rm K$ the temperature of the thermal background atoms
of density $n_b.$ 
Using $\sigma_{\rm Rb-Rb}=2.5\times 10^{-13}~\rm cm^2$\cite{ref.bali99}    
we find $\tau_c=55~\rm s$ at a pressure of $1\times 10^{-10}~\rm mbar.$ Even without resorting to cryogenic vacuum systems pressures as low as $10^{-11}~\rm mbar$ are achievable, which would imply collisional lifetimes of order 10 minutes. These estimates are consistent with observations\cite{ref.meschedesingleatom} of  FORT decay times using Cs atoms of 50 sec. at pressures of about $10^{-10}~\rm mbar$.

\subsection{Photon scattering}
\label{sec.trapscattering}

Scattering of FORT light by the qubit atoms causes some heating and leads
to a small amount of decoherence.  The scattering can be separated into
two contributions.  Elastic or Rayleigh scattering of the FORT light does
not change or dephase the qubit spin but does heat the external degrees
of freedom of the atoms.  Inelastic or Raman scattering occurs at a much
reduced rate but, since it changes  the spin-state of the qubit atom it
does contribute to decoherence, albeit at a small rate.

The elastic scattering cross section is 
\begin{equation}
\sigma_e=\frac{8\pi}{3} k_f^4 |\alpha_0|^2=1.6\times10^{-24}\;{\rm cm}^2
\end{equation}
where $k_f=2\pi/\lambda_f.$ 
To get the numerical value we have used the parameters given in Table \ref{tab.groundstate}.
  This scattering
 produces a heating rate
\begin{equation}
\frac{dE}{dt}=\frac{\hbar^2 k_f^2}{ 2 m}
\frac{\sigma_e I_f}{\hbar\omega_f}=\frac{\hbar\omega_f\sigma_e I_f}{ 2 m c^2}.
\end{equation}
Since both the heating rate and the trap depth $U$ scale with intensity,
their ratio gives the characteristic heating time for an atom in the
FORT
\begin{equation}
\frac{|U_m|}{ dE/dt}=\frac{2 \alpha_0 m\lambda_f}{ \hbar\sigma_e}=1950 \;{\rm
s}.
\end{equation}
A more conservative definition of the $T1$ due to Rayleigh scattering is the time for the atom to double its motional energy, which gives $T1=97~\rm s$ for the parameters of Table \ref{tab.groundstate}.

The inelastic scattering cross section can be expressed in terms of the vector polarizability $\alpha_1$ as 
\begin{equation}
\sigma_{\rm in}=\frac{4\pi}{3} k^4 |\alpha_1|^2=2.3\times10^{-27}\;{\rm
cm}^2
\end{equation}
where we have used $\alpha_1\approx -6.~\AA^3$.
The smallness of the inelastic cross section comes from the small coupling of the 
FORT light to
the electron spin of the atom which scales with the ratio of the 
fine-structure
splitting of the Rb P-levels to the detuning of the FORT laser.  

Since the inelastic scattering destroys the qubit state, it is a 
 source of decoherence.  It is proportional to FORT intensity, so it
can be reduced if necessary by operating at low trap depths.   The  qubit
longitudinal relaxation time due to inelastic scattering is

\begin{eqnarray}
T1_{\rm in}&=&\frac{1}{\sigma_{\rm ie}I_f/(\hbar\omega_f)} =\frac{3\hbar\lambda_f^3|\alpha_0|}{16\pi^3 |U_m|\alpha_1^2}\nonumber\\
&=& \frac {1}{|U_m|[{\rm mK}]}\times 151 \;{\rm s},
\label{eq.T1raman}
\end{eqnarray}
which is very long even for a  robust 1 mK trap depth.

\subsection{Laser noise induced heating}
\label{sec.traplasernoise}

Laser intensity and pointing fluctuations can cause undesirable heating
in FORTs\cite{Thomas97}.  The heating rate due to intensity noise is
\begin{equation}
\frac{dT_a}{ dt}={\pi^2}\nu^2 S_i(2\nu)T_a
\label{eq.heatrate1}
\end{equation}
where $\nu$ is the trap oscillation frequency, and
$S_i(2\nu)$ is the one-sided power spectrum of the fractional intensity
fluctuations.  These fluctuations are usually far above the shot-noise
limit  $S_i=2h\nu/P=4\times10^{-18}/{\rm Hz}$ at the 1-100 kHz frequencies
of interest here. As indicated above a 1 mK FORT depth requires 60 mW of $\lambda_f=1.01~\mu\rm m$ laser power 
which can be readily supplied by a small diode laser. A typical fluctuation level for an unstabilized 
diode
laser is    $S_i=10^{-12}/{\rm Hz}.$
The characteristic time for an atom to be heated out of the trap is 
\begin{equation}
T1=\frac{1}{\pi^2 \nu^2 S_i(2\nu)}\ln\left(\frac{|U_m|}{T_a}\right).
\end{equation}
Using the values given in Table \ref{tab.groundstate} and 
$\nu=39~\rm kHz$ gives $T1\simeq 50~ \rm s$.
  If necessary, feedback can be used to reduce the laser noise and extend the heating time. 

Fluctuations in the laser beam position are also a source of heating. The characteristic heating time can be written as 
\begin{equation}
T1 = \frac{ \langle x_a^2\rangle}{\pi^2 \nu^2 S_x(\nu)}\ln\left(\frac{|U_m|}{T_a}\right)
\end{equation}
where  $S_x$ is the frequency spectrum of the position fluctuations.
The parameters of Table  \ref{tab.groundstate} and Eq. (\ref{eq.xsq})
give $\sqrt{\langle x_a^2\rangle}=0.28~\mu\rm m. $ To obtain 
$T1=50~\rm s$ requires $\sqrt{S_x}=5.6\times 10^{-7}~\mu\rm m/\sqrt{Hz}$
which is feasible with careful attention to mechanical connstruction.

Finally for an anisotropic trap  there is also heating from beam-steering
fluctuations.  This implies, for a highly anisotropic trap of aspect
ratio  $\xi_f$, a heating time of 
\begin{equation}
T1 = \frac{1}{\pi^2 \nu^2 \xi_f^2 S_\theta(\nu)}\ln\left(\frac{|U_m|}{T_a}\right)
\end{equation}
where $S_\theta$ is the spectrum of angular fluctuations of the FORT laser beam. For the parameters we are using $\xi_f\sim 8$ so there is a strong sensitivity to beam-steering noise. 
Nonetheless it should be feasible using fiber optic delivery of the trapping beam to achieve $T1\sim 50~\rm s.$ 

To summarize this section, estimates of storage times due to technical noise induced heating are of order 50 s, for three different mechanisms. 
Without appealing to extraordinary technical developments we can set the total contribution due to technical laser noise as $T1\sim20~\rm s.$
Ultimately this number could be improved by several orders of magnitude before reaching limits set by quantum fluctuations.

\subsection{AC Stark shifts}
\label{sec.acstarkshift}

In the preceding sections we have discussed decoherence mechanisms that to an excellent approximation affect the qubit ground states equally. 
Therefore no dephasing of the qubit basis states is incurred and there is no contribution to a finite transverse relaxation time, $T2.$ 
As will be discussed in Sec. \ref{sec.qubit1} we will use the states $|a\rangle=|F=1,m_F=0\rangle$ and $|b\rangle= |F=2,m_F=0\rangle$
as our computational basis. In the absence of any applied fields these states have a hyperfine splitting $ U_{\rm hf}(0)/h\simeq6.83~\rm GHz.$ In the presence of a static electric field there is a correction to the hyperfine splitting in $^{87}$Rb given by\cite{ref.anderson61}
\begin{widetext}
\begin{equation}
 U_{hf}(E) \simeq  U_{\rm hf}(0)\left\{1-\alpha_s(0)  E^2 \left[\frac{1}{2(U_{n=5,L=1}-U_{n=5,L=0})}
+\frac{1}{U_{n=4,L=2}-U_{n=5,L=0}} \right]\right\}
\label{eq.wilmershift}
\end{equation}
\end{widetext}
where $\alpha_s(0)$ is the static polarizability and $E$ is the electric field amplitude. For a far detuned trapping laser with a photon energy that is small compared to the term differences that appear in Eq. (\ref{eq.wilmershift}) we can estimate the shift by making the replacement $\alpha_s(0)E^2\rightarrow \alpha_s(\omega_f)|{\mathcal E}|^2/2$ so that Eq.(\ref{eq.wilmershift}) can be written as
$U_{hf}({\mathcal E})=U_{hf}(0)-\beta U_{\rm ac}$ where $\beta = 2U_{hf}(0)[...],$ with $[...]$ the term in square brackets in (\ref{eq.wilmershift}). For $^{87}$Rb 
we find $\beta = 4.1 \times 10^{-5}.$ At a trap depth of $|U_m|=1~\rm mK$ 
the correction to $U_{\rm hf}$ is $\beta U_{\rm ac}\simeq 1500~\rm Hz.$

\begin{figure}[!b]
\includegraphics[width=7.cm]{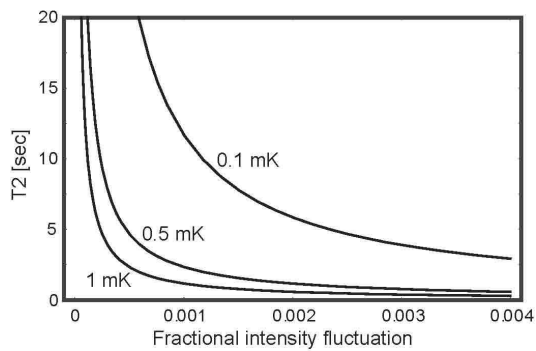}
\caption{Effective $T2$ due to differential AC stark shift of hyperfine levels for trap depths of $1,~ 0.5,$ and $0.1~\rm mK.$ }
\label{fig.ACstarkdecoherence}
\end{figure}

If the trapping laser had no intensity fluctuations and there was no atomic motion this small 
correction to the hyperfine shift would be time independent and would only give an unimportant correction to the hyperfine splitting. 
However, intensity noise and atomic motion result in a time dependent shift which gives a finite $T2.$ We consider first the effect of intensity noise which results in  state dependent dephasing due to fluctuations in $ U_{hf}({\mathcal E})$. When the averaged fluctuation vanishes over the time scale of interest there will be no additional decoherence, and indeed intensity fluctuations are not expected to be problematic
for   $\mu\rm s$  time scale gate operations.  However as regards storage of quantum information during a long calculation it is necessary to consider slow drifts in laser intensity that will give qubit dephasing.  We define an effective $T2$
due to dephasing  by 
\begin{equation}
T2=\frac{2\pi\hbar}{\delta U_{\rm hf}({\mathcal E})}=\frac{2\pi \hbar}{\beta U_{ac}}\frac{ I}{\delta I},
\label{eq.acstarkT2}
\end{equation}
where $\delta I/I$ is the fractional intensity fluctuation. 
The relative intensity noise is a function of frequency. In an actively stabilized system the fluctuations will be very small at high
frequencies. We are most concerned about finite fluctuations on the time scale of tens of seconds corresponding to the 
effective $T1$ given in Table \ref{tab.groundstate}.

The effective $T2$ is shown in Fig. \ref{fig.ACstarkdecoherence}
as a function of the fractional laser intensity fluctuation. At  a FORT depth of $|U_m|=1 ~\rm mK$ and a relative intensity fluctuation of $10^{-4},$ which is  feasible with active stabilization and well above the limit  set by quantum noise for mW power FORT beams, $T2\sim 12~\rm s.$

Atomic motion within the FORT volume leads to a time varying  trapping potential, and hence dephasing of the qubit states. This problem also arises in the context of precision measurements of optically trapped atoms\cite{ref.romalis,ref.chu}. 
A rough estimate says that since the atomic position spread is approximately $(T_a/|U_m|)^{1/2}w_{f0}/2,$ the fractional variation in trapping intensity due to the transverse motion is $\sim T_a/2|U_m|\sim 0.025$ for the parameters of Table \ref{tab.groundstate}.
The same fractional variation is also found for the axial motion.  This implies a motional 
variation in the hyperfine splitting of order $38 ~\rm Hz.$ The maximum phase perturbation in one axial vibrational period is thus $\sim 2\pi \times 38~{\rm Hz}/3~ {\rm kHz} \sim 
0.08~\rm rad.$   We can also express this shift as an effective transverse relaxation time $T2\sim  1/38 ~{\rm Hz}=.026 ~\rm s.$

This time is far shorter than the $T1$ and $T2$ times  due to the other mechanisms discussed above. As has been demonstrated experimentally in Ref.\cite{ref.davidson} it is possible to cancel the differential AC Stark shift of the hyperfine states by introduction of a weak beam tuned between the hyperfine states that has the same spatial profile as the FORT beam. The intensity of the additional compensation beam can be very low  such that the decoherence rates due to photon scattering will not change significantly. Kaplan, et al. \cite{ref.davidson} demonstrated a reduction in transverse broadening by a factor of 50, and we assume that a factor of at least 100 is realistic, in order to arrive at the estimate of $2.6 ~\rm s.$ given in Table \ref{tab.groundstate}.
Additional  discussion of motional effects in the context of single qubit operations is given in Sec. \ref{sec.qubit1}.

\subsection{Background magnetic and electric fields}
\label{sec.backgroundfields}

\begin{figure}[!t]
\includegraphics[width=7.5cm]{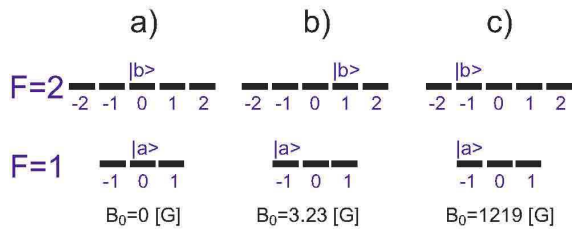}
\caption{Possible choices of qubit basis states that have a magic bias field  $B_0$ for which the relative 
Zeeman shift is a quadratic function of 
field fluctuations. 
 }
\label{fig.basisstates}
\end{figure}

The amount of dephasing caused by trapping and background field fluctuations depends on the qubit basis states that are used\cite{ref.wineland,ref.romalis}. 
As shown in Fig. \ref{fig.basisstates} for atoms with nuclear spin $I=3/2$ there 
are three possible choices of basis states that  are first order free of Zeeman shifts. 
In this subsection we consider the three possible choices and will conclude that set a) is optimal for achieving long storage times with low
decoherence.

The simplest choice (Fig. \ref{fig.basisstates}a) is to use $|a\rangle=|F=1,m_F=0\rangle$ and $|b\rangle=|F=2,m_F=0\rangle.$ 
Using the Breit-Rabi formula the second order shift of the energy interval expressed as a frequency is
\begin{equation}
\omega_{ba}=\omega_{\rm hf}\left[1+ \left(\frac{(g_S-g_I)\mu_B \delta B}{\hbar\omega_{\rm hf}}\right)^2\right]^{1/2}
\label{eq.zee2}
\end{equation}
where $\omega_{\rm hf}$ is the zero field hyperfine clock frequency between the $m_F=0$ states, $\mu_B$ is the Bohr magneton, $g_S, g_I$ are the electron spin and nuclear Land\'{e} factors,  
and $\delta B$ is the magnetic field fluctuation. The frequency deviation $\delta\omega_{ba}=\omega_{ba}-\omega_{\rm hf}$ implies a transverse relaxation time  $T2=2\pi/\delta\omega_{ba}$
which evaluates to   $T2=1740~\rm s$ for $\delta B=1~\rm mG.$ While it is in principle possible to shield magnetic field fluctuations to an even 
lower level it is difficult to do so in an experiment that requires substantial optical access to the atom trapping region. Recent work\cite{ref.bcontrol} has demonstrated 
suppression of static and fluctuating magnetic fields to the level of $300~\mu\rm G$ using an active feedback scheme. We will assume $1~\rm mG$ as a conservative estimate of the 
fluctuation level that can be achieved.

Alternatively we can use the states shown in Fig. \ref{fig.basisstates}b: $|a\rangle=|F=1,m_F=-1\rangle$ and $|b\rangle=|F=2,m_F=1\rangle.$ At a bias field of $B_0\simeq3.23~\rm G$ the frequency separation is quadratically dependent on fluctuations about $B_0.$ A fluctuation of 1 mG about the bias point gives $T2=2320~\rm s$ and 
 frequency separations between the qubit basis states and neighboring Zeeman states of about  2.3 MHz.

Finally  we could also use the states shown in Fig. \ref{fig.basisstates}c: $|a\rangle=|F=1,m_F=-1\rangle$ and $|b\rangle=|F=2,m_F=-1\rangle.$ At a bias field of $B_0=\hbar\omega_{\rm hf}/(2(g_S-g_I)\mu_B)$ the frequency separation is quadratically dependent on fluctuations about $B_0.$ Defining $\delta B=B-B_0$ we have 
\begin{equation}
\omega_{ba}=\frac{\sqrt 3}{2}\omega_{\rm hf}\left[1+ \frac{4}{3}\left(\frac{(g_S-g_I)\mu_B \delta B}{\hbar\omega_{\rm hf}}\right)^2\right]^{1/2}
\end{equation}
Apart from a factor of $2/\sqrt3$ larger sensitivity to fluctuations we retain the quadratic dependence of Eq. (\ref{eq.zee2})
at a very large bias field. For $^{87}$Rb we find $B_0\simeq1219.3~\rm G$ and $T2=1500~\rm s$ for $\delta B=1~\rm mG.$ At this large bias field the separation between neighboring $m_F$ levels is hundreds of MHz. This large detuning will effectively suppress unwanted transitions during logic operations, 
but has the disadvantage of mixing the hyperfine states so that there no longer will be a clean cycling transition between the $^2S_{1/2}|F=2,m_F=2\rangle$ and $^2P_{3/2}|F'=3,m_{F'}=3\rangle$ states that can be used for qubit measurement. One possibility would be to use a bias field that is applied during logic operations, and turned off adiabatically for state measurements.

The use of $m_F\ne 0$ basis states requires that we account for $m_F$ dependent shifts due to the ground state vector 
polarizability that couples to nonzero ellipticity of the trapping laser\cite{ref.happermathur,ref.cho,ref.romalis}. The energy shift can be written as
\begin{equation}
U_{\rm ac}^{(1)}=-\frac{1}{4}|\mcE_f|^2 \alpha_1 g_F m_F \sqrt{1-\epsilon^2} 
\label{eq.acstark2}
\end{equation}
where $g_F= [F(F+1)+S(S+1)-I(I+1)]/[F(F+1)]$, $m_F$ is the total spin projection along the FORT beam propagation direction $\hat y,$ and the laser polarization is $\bfepsilon=(1/\sqrt2)(\hat x\sqrt{1+\epsilon}+i\hat y \sqrt{1-\epsilon}).$ Basis states with $m_F=0$ have no vector shift, and in addition the field insensitive states $|0\rangle=|F=1,m_F=-1\rangle$ and $|1\rangle=|F=2,m_F=1\rangle$ both have 
$g_F m_F = 1/2$ so there is no differential shift and no decoherence due to the vector polarizability.
On the other hand the choice  $|0\rangle=|F=1,m_F=-1\rangle$ and $|1\rangle=|F=2,m_F=-1\rangle$ 
leads to a transverse decoherence time of $T2=(\alpha_0/\alpha_1)(I/\delta I) 2\pi \hbar/(k_B T_m \sqrt{1-\epsilon^2}).$
Using the parameters given in the caption of Table \ref{tab.groundstate}, $\epsilon=.999$ and a fractional intensity fluctuation of $\delta I/I=0.001$ we find  a short coherence time of $T2=38~\rm ms.$ We see 
that it is important to avoid decoherence  due to the vector polarizability so that the preferred choice is the 
qubit states shown in Fig. \ref{fig.basisstates}a or \ref{fig.basisstates}b.

Careful analysis along the lines of that used in Sec. \ref{sec.qubit1} shows that the set of Fig. \ref{fig.basisstates}b presents significant 
obstacles to achieving  high fidelity single qubit operations using stimulated two-photon Raman transitions. The essential problem is that the choice of  Fig. \ref{fig.basisstates}b involves
ground states separated by $|\Delta m_F|=2.$ For large detunings the two-photon Raman rate for these transitions is proportional to the vector polarizability which has a selection 
rule $|\Delta m_F|\le 1.$ The Raman rate therefore vanishes so it is necessary to use two Raman beams with opposite helicities. In this situation the effective Raman rate 
scales as $\Delta_e/\Delta_1^2$ where $\Delta_e$ is the width of the excited state hyperfine structure, and $\Delta_1$ is the one-photon detuning of the Raman beams. 
However, the Raman beam induced ac Stark shifts scale as $1/\Delta_1$ so that in the limit of large detuning the ground state ac Stark shifts become large  compared to the 
Raman rate between ground states. It is not possible, without resorting to more complex polarization states, to balance the ac Stark shifts of the qubit basis states, which leads to entanglement of the qubit spin state with the atomic center of mass motion. This entanglement represents an undesired decoherence mechanism.

We are thus led to the choice shown in Fig. \ref{fig.basisstates}a for the qubit basis states. 
While the $m_F=0$ states are insensitive to magnetic fluctuations they are not optimal as qubit basis states at low magnetic fields. Atomic motion in a region of near zero magnetic field is subject to Majorana transitions between Zeeman sublevels. 
Transitions can be suppressed by Zeeman shifting the states with a bias magnetic field. 
Unfortunately a large bias field converts the small field quadratic dependence of Eq. (\ref{eq.zee2}) to a linear dependence on the field fluctuations about the bias point. 
For example a bias field of 1 G, which gives MHz scale Zeeman shifts, with a fluctuation of 1 mG about the bias point would give $T2=0.87~\rm s$. We can do considerably better with  
a small $15~\rm mG$ bias field, which is sufficient to suppress Majorana transitions, yet small enough such that with $1~\rm mG$ of field fluctuations 
the coherence time is $T2=56~\rm s.$ In the remainder of this paper we will analyze the implementation of  quantum logic using  the $m_F=0$ basis states.

Finally we note that dephasing due to dc electric fields is completely negligible. The $T2$ due to differential ac Stark shifts calculated from Eq. (\ref{eq.acstarkT2}) results from a peak electric field at the center of a mK deep optical FORT of $O(10^6~\rm [V/m]).$  Since we expect low frequency field fluctuations to be much less than 1 V/m,  the dc Stark shift can be neglected.

\section{Single qubit operations}
\label{sec.qubit1}

A two-site FORT with a single atom loaded in each site provides a setting for studying basic one and two qubit operations. In this section we start with a study of  the fidelity and decoherence properties of one qubit operations at each site. Of particular concern will be the  requirement of high fidelity operations at a targeted site without unintended disturbance of the neighboring site. Simply increasing the separation of the sites to reduce crosstalk will imply slow 2-qubit conditional operations, so there is inevitably a performance trade-off between 1- and 2- qubit gates. We discuss balancing the conflicting requirements   in Sec.
\ref{sec.discussion} below.

\begin{table}[!t]
\centering
\begin{tabular}{l|c||c|c}
\hline
mechanism & Section & $p_{\rm dch}$& Error\\
\hline
spontaneous emission & \ref{sec.qubit1spe} & $9 \times 10^{-5}$
 &   \\
AC stark shifts & \ref{sec.acstarkshifts}& &$ 4.4\times 10^{-7}$\\
atomic motion& \ref{sec.atomiclocalization}& &$9.4\times 10^{-5}$\\
spatial crosstalk & \ref{sec.atomiclocalization}& &$2.2\times 10^{-5}$\\
polarization leakage & \ref{sec.qubitpolarization}&$9.7\times 10^{-5}$&\\
laser intensity noise& \ref{sec.qubit1lasernoise}&&$9.4\times 10^{-8}$ \\
laser phase noise &\ref{sec.qubit1lasernoise}&&$2.5\times 10^{-7}$\\
\hline
Combined & &$1.9\times 10^{-4}$& $1.2 \times 10^{-4}$\\
\hline
\end{tabular}
\caption{Physical mechanisms contributing to imperfection of single qubit operations. The decoherence probability is calculated for a $\pi$ rotation, and the fidelity error is calculated with respect to an ideal $\pi/2$ rotation using the metric of Eq. (\ref{eq.fidelity}). Values listed are for 
$T_m=1 ~\rm mK,$ 
$T_a=50~\mu\rm K,$ $\lambda_f=1.01~\mu\rm m,$ $w_{f0}=2.5~\mu\rm m,$ and $w_0=5~\mu\rm m.$  See text for details.}
\label{tab.singlequbit}
\end{table}

Single qubit rotations between ground state levels can be performed in several ways. Microwave fields that are resonant with $\omega_{ba}$  can be used, but do not allow direct single site addressing. By combining a microwave field with an electric or magnetic field gradient a selected site can be tuned into resonance\cite{ref.meschederegister}. The drawback of such an approach is that neighboring sites will be subjected to off-resonant perturbations.

Here we analyze an alternative approach using  stimulated two-photon Raman transitions induced by tightly focused addressing beams, as shown in Fig. \ref{fig.qubitrotation}. Two-photon Raman techniques for laser cooling of neutral atoms were pioneered by Kasevich and Chu\cite{ref.churaman}, and have been used recently in optical lattices by the group of Jessen\cite{ref.jessen} and others\cite{ref.raman}. Raman techniques are also an important ingredient  in trapped ion experiments\cite{ref.wineland}. 
As the physics of coherent state manipulation with stimulated Raman pulses is well understood our aim here is to analyze a number of contributions to nonideal behavior that arise in the context of optically trapped atom experiments. The physical mechanisms contributing to non ideal single qubit operations are summarized in Table \ref{tab.singlequbit} and discussed in Sections \ref{sec.qubit1spe}-\ref{sec.qubit1lasernoise}. In addition to qubit rotations fast state measurements are a requirement for error correction in a quantum processor. In section 
\ref{sec.qubit1fluorescence} we examine the speed and fidelity of single site state measurement using resonance fluorescence.

\subsection{Speed and decoherence}
\label{sec.qubit1spe}

As discussed in Sec.  \ref{sec.singleatomtraps} the  qubit logical basis states  $|a\rangle, |b\rangle$ 
will be represented with  the \Rb $5S_{1/2}$ $|F=1,m_F=0\rangle$ and 
$|F=2,m_F=0\rangle$    ground state hyperfine levels.  
Qubit initialization to the $|b\rangle$ state 
can be accomplished by driving the $5S_{1/2}(F=2) ~-~ 5P_{3/2}(F'=1)$ transition with a beam linearly polarized along the magnetic field. An additional repumper beam  returns atoms infrequrntly lost to the $F=1$ lower hyperfine manifold.   When the transition is driven with an intensity several times larger than the saturation intensity population will accumulate in  
$|F=2,m_F=0\rangle$ at the  rate of 
$\gamma/2$, where $\gamma/2\pi=5.98$ MHz is the spontaneous decay rate from the upper level. 
The characteristic qubit initialization time will be several transfer time constants, or  $\sim 0.1~\mu\rm s.$  
Experiments with  atomic beams\cite{ref.wiemanbeam} have demonstrated preparation purity by optical pumping at the level of $10^{-4}$.

\begin{figure}[!t]
\centering
\includegraphics[width=7.5cm]{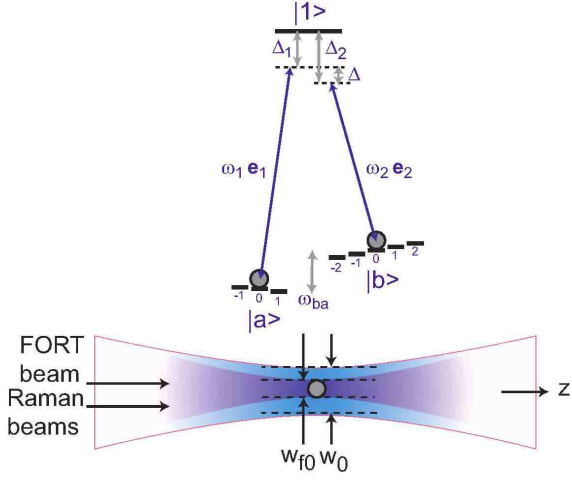}
\caption{(color online) Single qubit stimulated Raman rotations. }
\label{fig.qubitrotation}
\end{figure}

Ground state single qubit manipulations using stimulated Raman transitions can be performed with high fidelity and made quite free of decoherence due to 
 spontaneous emission. The driving fields and atomic level structure are shown in Fig. \ref{fig.qubitrotation}. We consider two driving fields at frequencies $\omega_1, \omega_2$ with detunings $\Delta_1=\omega_1-\omega_{1a}$, $\Delta_2=\omega_2-\omega_{1b}$, and associated Rabi frequencies $\Omega_{1,2}=d_{1,2}{\mathcal E}_{1,2}/\hbar$,  with $d_{1,2}$ the relevant dipole matrix elements between the ground states and the excited state $|1\rangle.$ The fields propagate along the $\hat z$ axis with 
polarizations $\bfepsilon_{1,2}$ so that the total optical field is 
\begin{widetext}
\begin{equation}
{\bf E} =  \frac{e^{-r^2/w^2(z)}}{\sqrt{1+z^2/z_R^2}}\left[
\frac{\mathcal E_1(t)}{2 }e^{\imath( k_1 z - \omega_1 t)} {\bfepsilon}_1
+ \frac{\mathcal E_2(t)}{2}e^{\imath (k_2 z - \omega_2 t)}{\bfepsilon}_2 \right]
+ c.c.
\end{equation}
\end{widetext}
where $w(z)=w_0\sqrt{1+z^2/z_R^2},$ 
 $z_R=\pi w_0^2/\lambda_R,$ $\lambda_R = \lambda_1,$ and we have taken the Rayleigh lengths of the two fields to be equal since $|(\lambda_1-\lambda_2)/(\lambda_1+\lambda_2)|\ll 1.$

Let us assume the atom is in state $|a\rangle$ at $t=0.$ The Raman light is tuned in the vicinity of either the $5P_{3/2}$ or $5P_{1/2}$ excited states. For definiteness we consider tuning near to the $5P_{1/2}$ excited state.  In the limit of large single photon detuning  relative to the excited state hyperfine structure the probability for the atom to be in state $|b\rangle$ at time $t$ is
\begin{equation}
|c_b(t)|^2=\frac{|\Omega_R|^2}{|\Omega_R|^2+\Delta^2}\sin^2\left(\frac{\sqrt{|\Omega_R|^2+\Delta^2} }{2}t \right)
\label{eq.pipulse}
\end{equation}
where  $\Omega_R=\Omega_1\Omega_2^*/\delta,$
$\delta=\Delta_1+\Delta_2 ,$ and
$\Delta=\omega_1-\omega_2-\omega_{ba}+(|\Omega_2|^2-|\Omega_1|^2)/(2\delta)
.$ When $|\Delta|\ll |\Omega_R|\ll |\Delta_1|,|\Delta_2|$ 
we have $\delta\simeq2\Delta_1$ and the effective Rabi frequency is $\Omega_R\simeq 
\Omega_1\Omega_2^*/(2\Delta_1).$ We assume that $\Delta_1,\Delta_2$ are small compared to the $^{87}$Rb fine structure splitting of 7120 GHz,
so we neglect any contribution from the other $5P$ state.

 The probability of spontaneous emission 
during  a $\pi$ pulse of time $t_\pi=\pi/\Omega',$ with $\Omega'=\sqrt{|\Omega_R|^2 + \Delta^2}$ the effective off-resonance Rabi frequency, is 
$p_{se}=\tau^{-1} \int_0^{t_\pi} dt  |c_p(t)|^2,$ where $c_p$ is the amplitude of the excited state and  $\tau=27.7~\rm ns$ is the $5$P$_{1/2}$ lifetime. Neglecting the two-photon detuning and assuming a piecewise constant pulse profile it is readily shown that $p_{se}\simeq \pi/(2|\Delta_1|\tau).$ For \Rb with $\Delta_1/2\pi=-100~\rm  GHz$
we get  $p_{se}=9 \times 10^{-5}.$

\begin{figure}[!t]
\centering
\includegraphics[width=6.cm]{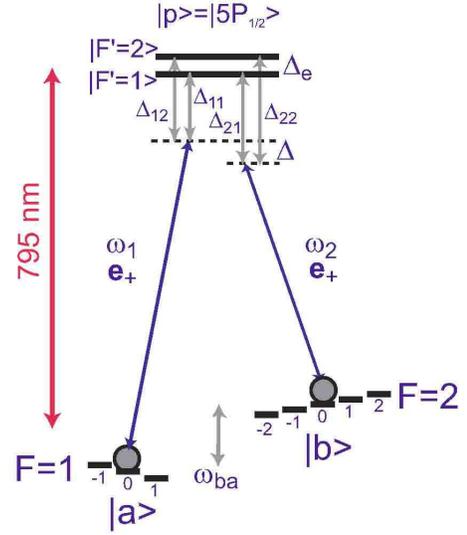}
\caption{(color online) Stimulated Raman rotations with two excited states. }
\label{fig.rabi2upper}
\end{figure}

The speed of two-photon Raman transitions scales with the optical intensity. In subsequent sections we will be concerned with 
corrections to the effective Raman frequency due to the excited state hyperfine structure shown in Fig. \ref{fig.rabi2upper}. We work in the limit of $|\Delta|\ll |\Omega_R|\ll |\Delta_{11}|,|\Delta_{12}|,|\Delta_{21}|,|\Delta_{22}|$, so the excited states are weakly populated.  With ${\bf e}_1={\bf e}_2={\bf e}_+$ the effective Rabi frequency is
\begin{equation}
|\Omega_R| = \frac{2 e^2 \left(R_{5S,5P}\right)^2}{\epsilon_0 c \hbar^2} K I \left| \frac{ \Delta_{11}-\Delta_e/4}{2\Delta_{11}(\Delta_{11}-\Delta_e)}\right|,
\label{eq.2upperrabi}
\end{equation} 
with $R_{5S,5P} \simeq5.13 a_0$ the radial integral between the 
ground and excited states, $a_0$ is the Bohr radius, $K=1/9$ is an angular factor, $I$ is the intensity of each Raman beam, and
  $\Delta_e/2\pi = 817~\rm  MHz$  is the hyperfine splitting of the  excited state.
  Working with $100~\mu\rm  W$ in each beam focused to a spot with waist $w_0=5~\mu\rm m$ at a detuning of $|\Delta_{11}|/2\pi = 100~\rm  GHz$ we get
$\Omega_R/2\pi = 4.6~\rm MHz$.

\subsection{Raman beam AC Stark shifts}
\label{sec.acstarkshifts}

In this section we calculate heating and decoherence effects due to  the tightly focused Raman beams.
Before proceeding with calculations of the fidelity of qubit operations we define the metric to be used.
Starting with an initial pure state $|\psi\rangle=c_a|a\rangle+c_b|b\rangle$ a two-photon stimulated Raman transition 
results in the transformation $|\psi\rangle\rightarrow R(\Omega_R,\Delta,t)|\psi\rangle$ where the rotation matrix is 
\begin{widetext}
\begin{equation}
R(\Omega_R,\Delta,t)=\left(
\begin{matrix}
e^{\imath \Delta t/2}\left[ \cos\left( \frac{\Omega' t}{2} \right) -i \frac{\Delta}{\Omega'}\sin\left( \frac{\Omega' t}{2} \right) \right] &
 i  e^{\imath \Delta t/2}\frac{\Omega_R^*}{\Omega'}\sin\left( \frac{\Omega' t}{2} \right) 
\cr 
ie^{-\imath \Delta t/2}  \frac{\Omega_R}{\Omega'}\sin\left( \frac{\Omega' t}{2} \right)&
e^{-\imath \Delta t/2}\left[ \cos\left( \frac{\Omega' t}{2} \right) +i \frac{\Delta}{\Omega'}\sin\left( \frac{\Omega' t}{2} \right) \right]
\end{matrix}
\right).
\label{eq.Rlong}
\end{equation}
\end{widetext} In writing Eq. (\ref{eq.Rlong}) we have suppressed a multiplicative phase factor and neglected a small correction to some of the terms containing $\Delta$ that is proportional to the differential  ac Stark shift of the basis states.
The fidelity of a rotation operation $R(\Omega_R,\Delta,t)$ compared to an ideal transformation
with  $R_0=R(\Omega_{R0},\Delta_0,t_0)$  can be defined as 
\begin{equation}
F= \left<|\langle R_0\psi | R\psi\rangle|^2 \right>,
\label{eq.fidelity}
\end{equation}
where the outer brackets specify an average over any stochastic contributions to $R.$

In the simplest case of  two-photon resonance $\Delta=0$ and the rotation matrix simplifies to 
\begin{equation}
R(\theta,\phi)=\left(\begin{matrix}\cos(\theta/2) & i e^{-\imath \phi}\sin(\theta/2)\\
 i e^{\imath \phi}\sin(\theta/2) & \cos(\theta/2)\end{matrix}\right)
\label{eq.Rsimple}
\end{equation}
where  $\theta=|\Omega_R|t$ and  $\phi=\arg(\Omega_R).$
For the particular case of an ideal  $\pi$ pulse $(\theta_0=\pi,\phi_0=0)$ with $|\psi\rangle=|a\rangle$ the fidelity is $F=(1/2)(1-\cos\theta).$ As this metric is independent of errors in the azimuthal angle $\phi$ 
we find it more informative to quantify the fidelity of a single operation with respect to an ideal $\pi/2$ pulse. In this case the fidelity is $F=(1/2)(1+\cos\phi\sin\theta).$ The gate errors listed in Table \ref{tab.singlequbit}
are defined by $E=1-F.$ The extent to which errors accumulate  in concatenated qubit operations is an important consideration when designing a computational sequence. A discussion of this topic in the context of an ion trap experiment has been given in \cite{ref.wineland}.

Returning to the effect of ac Stark shifts we note that in addition to providing controlled rotations between the qubit basis states the Raman beams result in  unequal ac Stark shifts of the Zeeman states
which leads to an additional rotation 
phase $\delta\phi.$
The rotation phase has an average value   
that must be accounted for\cite{ref.blattstarkshift}, as well as a stochastic part resulting from atomic motion that leads to a fidelity error. The Raman beam induced Stark shifts also play a useful role by enhancing the nondegeneracy of the Zeeman states beyond that provided by the very small bias magnetic field.  This effect greatly reduces  leakage out of the computational basis as we discuss in Sec. \ref{sec.qubitpolarization} below. 

\begin{widetext}

\begin{table}[!t]
\begin{tabular}{|l||c|c|c|}
\toprule
ground state & $K$ &  $\delta U_R ~\rm [mK]$  &  $[\delta U_R-\delta U_R(|a\rangle)]/h ~\rm [MHz]$  \\
\colrule
   $|1-1\rangle $& $\frac{1}{\Delta_{11}}+\frac{1}{\Delta_{11}-\Delta_e}+\frac{1}{\Delta_{11}-\omega_{ba}}+\frac{1}{\Delta_{11}-\Delta_e-\omega_{ba}} $  & $- 0.107$& $-2.41$ \\
    $ |10\rangle~(=|a\rangle)$ &$\frac{1}{\Delta_{11}}+\frac{3}{\Delta_{11}-\Delta_e}+\frac{1}{\Delta_{11}-\omega_{ba}}+\frac{3}{\Delta_{11}-\Delta_e-\omega_{ba}} $ &$ -0.213 $&$0$\\
    $ |11\rangle$ & $\frac{6}{\Delta_{11}-\Delta_e}+\frac{6}{\Delta_{11}-\Delta_e-\omega_{ba}} $ & $-0.319$&$2.42$\\
\colrule
 &  &    & $[\delta U_R-\delta U_R(|b\rangle)]/h ~\rm [MHz]$ \\
\colrule
     $ |2-2\rangle$ &$\frac{6}{\Delta_{11}+\omega_{ba}}+\frac{2}{\Delta_{11} -\Delta_e+\omega_{ba}}+\frac{6}{\Delta_{11}}+\frac{2}{\Delta_{11}-\Delta_e}$ & $-0.458$&-4.79\\    
 $ |2-1\rangle$ & $\frac{3}{\Delta_{11}+\omega_{ba}}+\frac{3}{\Delta_{11}-\Delta_e+\omega_{ba}}+\frac{3}{\Delta_{11}}+\frac{3}{\Delta_{11}-\Delta_e} $& $-0.343$&$-2.39$\\
     $ |20\rangle~(=|b\rangle)$& $\frac{1}{\Delta_{11}+\omega_{ba}}+\frac{3}{\Delta_{11}-\Delta_e+\omega_{ba}}+\frac{1}{\Delta_{11}}+\frac{3}{\Delta_{11}-\Delta_e} $ &$-0.228$ &$0$\\
 $ |21\rangle $ & $\frac{2}{\Delta_{11}-\Delta_e+\omega_{ba}} +\frac{2}{\Delta_{11}-\Delta_e} $ &$-0.114$&$2.38$\\
 $|22\rangle $ &$0$  &$0$&$4.75$\\
\botrule
\end{tabular}
\caption{Stark shifts of Zeeman ground states due to the Raman beams of Fig. \ref{fig.rabi2upper}.
The Stark shifts given by $\delta U_R= ( e^2 R_{5S,5P_{1/2}}^2/(\epsilon_0 c\hbar))(K/72)I$, with $I$ the intensity of each Raman beam, are evaluated in columns $3$ and $4$  for 
$I=100~\mu\rm W$ and $\Delta_{11}/2\pi=-100~\rm GHz.$ 
}
\label{tab.acramanstark}
\end{table}

\end{widetext}

In order to give a quantitative account of the ground state Stark shifts for Raman beams tuned near the D1 line we add the contributions from the $|F'=1,2\rangle$  excited states  for both beams to get the results shown in Table \ref{tab.acramanstark}. 
Using  the physical parameters given in the previous section we find the change in trapping potential at the center of one of the Raman beams is  $U_R\sim 220~\mu\rm K$ which is a few times less than the $|U_m|=1~\rm mK$ wells created by the FORT beams. Using red detuned Raman beams the additional potential is attractive, nonetheless the associated dipole forces can lead to heating of the trapped atoms. In the limit where the Rabi frequency is much larger than the trap oscillation frequency a trapped atom will only move a small fraction of its orbit during a 
Raman pulse. The dipole force during the pulse will with equal probability accelerate or decelerate the atomic motion, so that on average, to lowest order in the ratio of trap frequency to Rabi frequency, there will be no heating.  Another way to see this is to note that the heating rate 
given by Eq. (\ref{eq.heatrate1}) vanishes when the perturbation has no energy at twice the trap oscillation frequency. 

Nonetheless when we consider the effect of a single  Rabi pulse there will be a worst case heating of the atomic motion that we require to be small compared to the trap depth to avoid loss of the trapped atom. It is readily shown that the  heating due to a single Rabi $\pi$ pulse is bounded by 
$\delta U = 2 \pi U_R (w_{f0}^2/2 w_{0}^2) (T_a/|U_m|) (\omega_x/\Omega_R) .$
Using the parameters given in Tables \ref{tab.groundstate} and \ref{tab.singlequbit} and $\Omega_R=2\pi \times 4.6 ~\rm MHz,$
we find $\delta U \sim 70~\rm n K.$ Since this amount of heating is very small compared to the FORT depth, it will not lead to escape of the 
trapped atom.  Note that in the limit of fast Rabi frequency the amount of heating is proportional to the atomic temperature, but is independent of the laser intensity used for the Raman pulse. 
Although in the first approximation the heating per Rabi rotation averages to zero there is a contribution to the heating rate proportional to the square of the trap frequency. 
For  $T_a=50 ~\mu\rm  K$ we get  an average energy increase of $70~\rm n K$ per operation so the maximum number of operations 
before there is significant heating of the atoms is about $ 10^3.$ This implies that recooling of the atomic motion after state measurements will be necessary to enable 
many logical operations.

To quantify the fidelity error due to the  Raman beam induced ac Stark shifts we note that a Rabi rotation will result in a transformation $|\psi\rangle=c_a|a\rangle + c_b|b\rangle\rightarrow c_a'|a\rangle + e^{\imath \delta\phi}c_b'|b\rangle,$  where $c_a', c_b'$ are the desired 
result of the Rabi rotation, and $\delta\phi$ is an additional differential phase shift due to the ac Stark 
shifts. The differential phase can be written as  $\delta\phi=\delta\phi_b-\delta\phi_a=t (\delta U_b - \delta U_a)/\hbar $
with $t$ the length of the pulse.  
The Raman field $\mathcal E$ that is seen by the trapped atom is time dependent  due to the atomic motion at finite temperature. The time averaged differential phase is proportional to $\langle \delta \phi \rangle\sim \langle {\mathcal E}^2 \rangle$
and the variance of the phase shift is $\Delta^2(\delta \phi)= \langle \delta\phi^2\rangle - \langle \delta\phi \rangle^2.$ Accounting for the atomic motion in the two transverse dimensions, and neglecting the axial motion which gives a much smaller contribution to time variation of the field, we find
$\langle{\mathcal E}^2 \rangle= {\mathcal E}_0^2 /\left[1+ (T_a/|U_m|)(w_{f0}/w_0)^2 \right]$
and  
$\langle{\mathcal E}^4 \rangle= {\mathcal E}_0^4 /\left[1+ 2(T_a/|U_m|)(w_{f0}/w_0)^2 \right],$ with 
${\mathcal E}_0$ the peak value of the field. 
These expressions are valid in the limit of tight confinement where $T_a\ll |U_m|.$

Using Eq. (\ref{eq.2upperrabi}) and Table \ref{tab.acramanstark} we find to leading order in the ratio of the atomic temperature to the trap depth 
for a $\pi/2$ pulse of length $t_{\pi/2} = \pi/2\langle|\Omega_R| \rangle$
\begin{widetext}
\begin{subequations}
\begin{eqnarray}
\langle \delta\phi \rangle&=& -\frac{\pi}{2}\frac{ \omega_{ba}}{\Delta_{11}}-\frac{3\pi}{8}\frac{ \Delta_e\omega_{ba}}{\Delta_{11}^2}+O(\omega_{ba}^3/\Delta_{11}^3)
\label{eq.dphi}
\\
\Delta^2(\delta \phi)&=&
\left(\frac{T_a}{|U_m|}\right)^2 \left( \frac{w_{f0}}{w_0}\right)^4 
\left|\frac{\pi}{2}\frac{ \omega_{ba}}{\Delta_{11}}+\frac{3\pi}{8}\frac{ \Delta_e\omega_{ba}}{\Delta_{11}^2}+O(\omega_{ba}^3/\Delta_{11}^3) \right|^2
\label{eq.vardphi}
\end{eqnarray}
\end{subequations}
\end{widetext}
The average phase shift given by Eq. (\ref{eq.dphi}) evaluates to $0.1~\rm rad$ for $\Delta_{11}/2\pi=-100~\rm GHz.$
This phase can be compensated for by adjustment of the relative phase of the Raman beams. It turns out using the full dependence of differential phase on detuning that  there are finite values of the detuning such that the ac Stark shifts are equal and the differential phase vanishes\cite{ref.choshift}. However, these detunings are of order the hyperfine ground state splitting, and are too small to suppress spontaneous emission from the $5P_{1/2}$ states. 
The standard deviation of the stochastic phase given by Eq. (\ref{eq.vardphi}) scales  linearly with the factor  $(T_a/|U_m|)(w_{f0}/w_0)^2$ which expresses the amount of variation of the Raman intensity over the cross sectional area the atom is confined to.
Using Eq. (\ref{eq.fidelity}) the gate error is $E\simeq(1/4)\Delta^2(\delta \phi)=4.4\times 10^{-7}$ at -100 GHz detuning.

\subsection{Atomic position and velocity fluctuations}
\label{sec.atomiclocalization}

In addition to fluctuations in the phase of the Rabi rotation, 
variations in the atomic position and velocity also directly perturb the angle of single qubit rotations since the effective pulse area depends on the local value of the Raman beam intensities as well as motional detuning due to Doppler shifts. 
Starting with an atom in state $|a\rangle,$ and applying the Raman fields ${\mathcal E}_1, {\mathcal E}_2$  the probability for the atom to be rotated to state $|b\rangle$ after time $t$
is given by Eq. (\ref{eq.pipulse}). Fluctuations in the atomic position and momentum lead to fluctuations in the effective pulse area at time $t_\pi$. 

 We can characterize a $\pi/2$ pulse and its fluctuations due to atomic motion by
\begin{eqnarray}
\left<|\Omega_R| t_{\pi/2} \right> &=& \pi/2, \left<|\Omega_R|^2 t_{\pi/2}^2 \right> = \pi^2/4+
\left< \epsilon_1^2 \right>,\\
\left<\Delta t_{\pi/2} \right> &=& 0, ~~~~\left<\Delta^2 t_{\pi/2}^2 \right> =\left< \epsilon_2^2 \right>,
\end{eqnarray}
where we have assumed the system has been prepared with zero detuning and $\epsilon_1,\epsilon_2$ are small stochastic parameters. 
With the Raman beams copropagating the two-photon detuning is first order Doppler free. Taking account of the velocity spread given by Eq. (\ref{eq.trapmotion}) we find
\begin{equation} 
\left<\epsilon_2^2\right>=\frac{\pi^2 \omega_{ba}^2}{4c^2\Omega_R^2} \frac{T_a}{m}.
\label{eq.dopplererror}
\end{equation}
At $T_a=50~\mu\rm K$ we find $\left<\epsilon_2^2\right>\simeq 3 \times 10^{-13}$ so we can neglect the 
contribution of Doppler detuning to the rotation error and use the simplified rotation matrix of Eq. (\ref{eq.Rsimple})
with $\phi=0.$

Averaging over the atomic motion in the same way as in the previous section we find
\begin{equation} 
\left<\epsilon_1^2\right>=\frac{\pi^2}{4}\left(\frac{T_a}{T_m} \right)^2 \left(\frac{w_{f0}}{w_0} \right)^4 
\end{equation}
and a fidelity error of $E=(1/4)\left<\epsilon_1^2\right>.$ This result takes account of the two-dimensional transverse motion of the trapped atom. Adding in the axial motion gives an additional factor proportional to 
 $\xi_f^2/\xi^2$, where $\xi=\pi w_0/\lambda_R.$ Our standard system parameters give $\xi_f^2/\xi^2=0.15,$ so this is a small correction which we will neglect. With the parameters of Table \ref{tab.singlequbit} we find
$E=9.4\times 10^{-5}$ at a temperature of $T_a=50~\mu\rm K.$ This error is larger than that due to ac Stark shifts by a factor of $(\Delta_{11}/\omega_{ba})^2$, but is still very small at typical sub Doppler temperatures that are easily reached in a MOT.
 
As the motional error scales inversely with the waist of the Raman beams  it is desirable to use as large a waist as possible.  In a multiple site device unwanted crosstalk occurs if the waist is made comparable to the site to site spacing  $d$. Simply increasing $d$ is not feasible since that would  reduce the fidelity of two-qubit operations, as will be discussed in Sec. \ref{sec.qubit2}.
The application of Raman beams giving a Rabi frequency $\Omega_R$ at the addressed qubit will result in 
a leakage Rabi frequency at a neighboring qubit of $\Omega_R'=\Omega_R e^{-2d^2/w_0^2}.$
For parameters that give a $\pi/2$ rotation at the targeted site there will be a fidelity error of $E=(\pi^2/16) e^{-4 d^2/w_0^2}$ at the neigboring site. We will use a site spacing of $d=8~\mu\rm m$ which gives $E=2.2\times 10^{-5} $ for $w_0=5~\mu\rm m.$  In practice laser beams that pass through a large number of optical elements may deviate significantly from an ideal gaussian profile. Full characterization of spatial crosstalk will depend on specific experimental details.

\subsection{Polarization effects}
\label{sec.qubitpolarization}

Two-photon stimulated Raman transitions may result in the atom being transferred to  a Zeeman state that  
lies outside the computational qubit basis. This will occur when  the Raman beams are a mixture of polarization states. 
The connection between fidelity loss due to unwanted transitions and the  polarization impurity of the Raman beams is calculated in Appendix \ref{sec.appendixpolarization}. 
With careful attention to optical design we may achieve $\delta\sim 10^{-3}$ for a wide beam.
In order to control one qubit at a time the Raman beams are focused to a waist of $w_0=5~\mu\rm m.$
 Near the focus of a linearly polarized Gaussian beam propagating along $\hat z$ the positive frequency component 
of the field can be written as
${\bf E}(x,y,z,t)={\bf e}_x({\mathcal E}_0/2) e^{-(x^2+y^2)/w_0^2}e^{\imath(kz-\omega t)}.$ 
Consistency with Maxwell's equations requires that the actual field is
${\bf E}(x,y,z,t)=[{\bf e}_x-i(x/z_R){\bf e}_z]({\mathcal E}_0/2) e^{-(x^2+y^2)/w_0^2}e^{\imath(kz-\omega t)}$
which includes  a component of ${\bf e}_z$. Thus a circularly polarized field, ${\bf E}\sim {\bf e}_+$ becomes 
${\bf E}(x,y,z,t)=[{\bf e}_+ +((y-ix)/z_R){\bf e}_0]({\mathcal E}_0/2) e^{-(x^2+y^2)/w_0^2}e^{\imath(kz-\omega t)}.$
We can make a rough estimate of the magnitude of the polarization induced leakage by inserting into Eqs. (\ref{eqs.leaka},\ref{eqs.leakb}) 
$|\epsilon_{ij}|\simeq \sqrt{\langle x_a^2 \rangle}/z_R=[w_{f0}/(k w_0^2)]\sqrt{T_a/|U_m|}=0.0028,$
with the numerical value calculated for our standard parameters and $T_a=50~\mu\rm K.$
This estimate puts an upper limit on the effective polarization purity in the interaction region, 
even when the unfocused Raman beams are perfectly polarized.  Using this value for all coefficients $\epsilon$ gives the 
last column in Table \ref{tab.polarization} which shows that transition amplitudes to undesired states will not 
exceed $\sim 6.6 \times 10^{-3}.$  Leakage out of the computational basis is a source of decoherence. 
We characterize the decoherence probability for a $\pi$ pulse by adding the probabilities for 
leakage out of states $|a\rangle$ or $|b\rangle$.  
We find starting in  $|a\rangle$ a leakage probability  of $5.6\times 10^{-5}$, and starting in $|b\rangle$ 
$9.4\times 10^{-5}.$ We use the larger of these numbers in Table \ref{tab.singlequbit} as an estimate of the 
decoherence probability  due to polarization effects.

\subsection{Laser intensity noise and linewidth}
\label{sec.qubit1lasernoise}

Intensity fluctuations of the Raman lasers will impact the accuracy of the Rabi pulse area. 
With the average intensity and pulse length set to give a $\pi/2$ pulse, 
a relative fluctuation of $\delta I/I$ implies a fidelity error of 
$E=(1/4)(\delta I/I)^2 .$ 
Active stabilization of the Raman laser intensity is limited by shot noise to\cite{ref.wineland} 
$$
\frac{\delta I}{I}\ge \left(\frac{4\hbar \omega_1}{\eta P t_{\pi/2} } \right)^{1/2}
$$
where $P$ is the power of the Raman beam and $\eta$ is the quantum efficiency of the detector in the stabilization circuit. With $P=100~\mu\rm W$, $|\Omega_R|/2\pi=4.6\times 10^6~\rm Hz,$ and $\eta=0.5$
we find $\delta I/I\ge 10^{-4}$ and $E\ge 9.4 \times 10^{-8}.$

Finite laser linewidth, and in particular relative phase fluctuations of the two Raman beams 
will lead to errors in the phase $\phi$ of the qubit rotation.  Laser oscillators have been demonstrated with 
a fractional frequency instability of $3\times 10^{-16}$ at $1 ~\rm sec$ averaging time\cite{ref.bergquistlaser}.
An optical phase lock between two laser oscillators with a residual phase noise at the $1 ~\mu\rm rad$ level has 
also been achieved\cite{ref.hallmicroradian}. As a conservative estimate we will assume the Raman lasers can be prepared with a relative phase 
noise of $1 ~\rm m rad$. This implies a fidelity error for a $\pi/2$ rotation  of
$E=2.5\times 10^{-7}.$

\subsection{State detection using resonance fluorescence}
\label{sec.qubit1fluorescence}

Rapid state selective measurements can be made by illuminating the atom with ${\bf e}_+$ polarized light tuned close to the $|5S,F=2,m_F=2\rangle \leftrightarrow |5P,F=3,m_F=3\rangle$ cycling transition. 
State readout can be based on detection of  resonance fluorescence\cite{ref.winelandshelving}. Alternatively  amplitude\cite{ref.winelandfluorescence}  and/or phase shifts imparted to a tightly focused probe beam can be used. In either case Poissonian photon counting statistics result in measurement times that are several orders of magnitude longer than Rydberg gate operation times. While the use of subPoissonian light could be advantageous in this context, it would add additional complexity. 

It is of interest to estimate the time for performing a state measurement with a desired accuracy.  The number of photon counts recorded in a measurement time $\tau$ is 
\begin{equation}
q(\tau)= \eta \frac{\Omega_d}{4\pi}\frac{\gamma \tau}{2}\frac{I/I_s}{1+4\frac{\Delta^2}{\gamma^2}+I/I_s}
\end{equation}
where $\Omega_d$ is the solid angle of the collection optics, $\gamma$ is the radiative linewidth, $I$ is the readout intensity, $I_s$ is the saturation intensity, and $\Delta$ is the detuning of the readout light from the cycling transition. The factor $\eta<1$ accounts for the quantum efficiency of the detector as well as any optical losses. The counts are assumed poisson distributed so that the probability of measuring $n$ counts is
$P_n(q) = e^{-q}q^n/n!.$ We also assume a background count rate $b_0$ due mainly to parasitic scattering from optical components and detector dark counts that gives a count number $b(\tau)= b_0 \tau.$

To make a measurement we detect scattered photons  for a time $\tau$. If the number of counts is greater than or equal to a cutoff number $n_c$ we have measured the qubit to be in state $|b\rangle.$ The measurement is incorrect if the actual number of signal counts was less than $n_c$ or the number of background counts was greater than or equal to $n_c.$ The probability of a measurement error is therefore 
\begin{eqnarray}
E &=& \sum_{n=0}^{n_c-1} P_n(q) + \sum_{n=n_c}^\infty P_n(b)\nonumber\\
&=& 1- \frac{\Gamma(n_c,b)}{\Gamma(n_c)} + \frac{\Gamma(n_c,q)}{\Gamma(n_c)},
\end{eqnarray}
where $\Gamma(n_c)= (n_c-1)!$ and $\Gamma(n_c,q)=\int_q^\infty dt~t^{n_c-1} e^{-t}$ is the incomplete gamma function. Since $\Gamma(n_c,0)=\Gamma(n_c)$ the error vanishes when $b\rightarrow 0$ and $q\rightarrow \infty$. 
For given values of $q$ and $b$ there is an optimum choice of $n_c$ that minimizes the error.

\begin{figure}[!t]
\centering
\includegraphics[width=7.5cm]{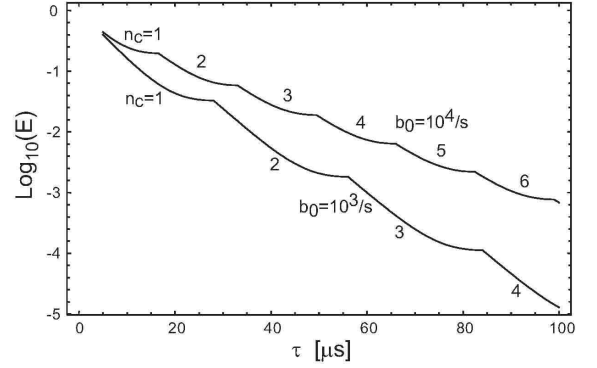}
\caption{State detection by resonance fluorescence for two different background count rates. The curves are labelled with the optimum values of $n_c$.  Parameters used were 
$\Omega_d/4\pi=.05,$ $\Delta/\gamma=-1/2,$ $I/I_s=1,$ $\gamma=2\pi \times 6  ~\rm MHz,$ and $\eta=0.6.$ }
\label{fig.statedetection}
\end{figure}

Figure \ref{fig.statedetection} shows the error probability as a function of measurement time for experimentally realistic parameters. We see that optimum detection corresponds to a very small value for $n_c$  and that  even with background rates as
high as $b_0=10^4 ~\rm s^{-1}$ accurate measurements can be made in under $100~\mu\rm s.$ A problematic aspect of the state measurement process is  concomitant heating of the atomic motion. The calculations shown assume a detuning of $\Delta=-\gamma/2$, so that if  counterpropagating readout beams are used it should be possible to cool the atomic motion while performing the measurement. Experiments have demonstrated the feasibility of long measurement times exceeding several seconds for single atoms confined in micron sized optical traps\cite{ref.grangiernature,ref.meschedetransport}.

\section{Two-qubit phase gate}
\label{sec.qubit2}

\begin{figure}[!b]
\centering
\includegraphics[width=7.5cm]{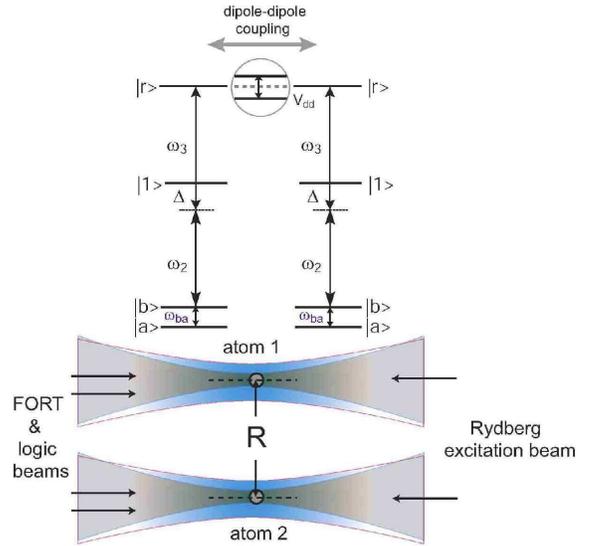}
\caption{(color online) Geometry of two-qubit interactions. }
\label{fig.doubleqgeometry}
\end{figure}

In this section we study the performance of the Rydberg gate using the  geometry of two trapped atoms separated by a distance $R$, as shown 
in Fig. \ref{fig.doubleqgeometry}.  The basic idea of the Rydberg gate is to use the strong dipole-dipole interaction of highly excited atoms to give a fast conditional phase shift. 
The fidelity of a conditional two-qubit operation will be impacted by the mechanisms affecting single qubit gates listed in Table \ref{tab.singlequbit}, plus  additional effects specific to the use of Rydberg states. 
As was shown above the single qubit imperfections can be quite small, leading to projected fidelity errors 
$O(10^{-4})$. In the following subsections  we analyze the additional errors specific to a conditional phase gate. 

In comparison to the performance of single qubit operations there are two significant complications involved in achieving conditional logic. The first is that optical excitation of Rydberg levels  cannot readily be made Doppler free so that atomic motion introduces pulse area errors. The second, and more serious limitation, arises from motional heating due to transfer to Rydberg states, and   
decoherence due to the finite lifetime of the Rydberg states. We present a solution to the heating problem based on balancing ground and Rydberg state polarizabilities, and show that the decoherence rates can be managed such that high fidelity gate operations appear possible. 

The Rydberg gate can be optimized in two limits. In the first limit (Sec. \ref{sec.qubit2rabilarge}), where the atoms are relatively far apart, the two-atom interaction frequency shift is small compared to the Rabi frequency of the Rydberg state excitation. 
In the opposite limit of closely spaced atoms (Sec. \ref{sec.qubit2ddlarge})
the dipole-dipole interaction is large compared to the Rabi frequency.
In both cases the finite lifetime of the Rydberg state sets a lower limit on the gate fidelity that scales as 
the smaller of 
$(\Omega_R \tau)^{-2/3}$ or $(\Delta_{dd} \tau)^{-2/3}$, with $\Omega_R$ the Rabi frequency of the Rydberg excitation, $\Delta_{dd}$ the dipole-dipole interaction shift, and $\tau$ the excited state lifetime.

We start the analysis of the phase gate by calculating its intrinsic fidelity scaling with speed of operation. 
The dipole-dipole interaction energy is
\begin{equation}
V_{dd}=\hbar\Delta_{dd}=\frac{1}{4\pi\epsilon_0 R^3}\left[\boldsymbol \mu_1\cdot\boldsymbol\mu_2 - 3
\frac{(\boldsymbol\mu_1\cdot {\bf r}_{12}) (\boldsymbol\mu_2\cdot {\bf r}_{12}) }{R^2}\right].
\label{eq.deltadd}
\end{equation}
Here $\boldsymbol\mu_j$ is the dipole moment of atom $j,$ and ${\bf r}_{12}=R \hat{\bf r}_{12}$ is the atomic separation. 
When the site to site qubit spacing $R$  is sufficiently large  the dipole-dipole interaction shift $\Delta_{dd}$ is a small perturbation compared to $\Omega_R$, the Rabi frequency for excitation from $|b\rangle\rightarrow |r\rangle.$
The protocol for a phase gate in this limit is\cite{ref.cote}: i) excite both atoms from $|b\rangle\rightarrow |r\rangle$ with a $\pi$ pulse, ii) wait a time $t_{dd}=\pi/\Delta_{dd}(R),$ and iii) transfer both atoms back down from $|r\rangle\rightarrow |b\rangle$ with a $\pi$ pulse. Here $\Delta_{dd}(R)$ is the dipole-dipole shift at an atomic spacing of $R.$ The resulting idealized logic table is: $|aa\rangle\rightarrow|aa\rangle,$ 
$|ab\rangle\rightarrow-|ab\rangle,$
$|ba\rangle\rightarrow-|ba\rangle,$
$|bb\rangle\rightarrow-|bb\rangle$ which is an  entangling phase gate. Adding single qubit Hadamards before and after the conditional interaction implements a CNOT gate. 
The fidelity of the gate is constrained by the presence of four time scales. In the large Rabi frequency limit we have $1/\omega_{ba}\ll  1/|\Omega_R|\ll 1/\Delta_{dd}\ll \tau$, where $\tau$ is the natural lifetime of the Rydberg state.  The fidelity is fundamentally limited by the combination 
$\omega_{ba}\tau$ which should be as large as possible. Using characteristic values of $\omega_{ba}=2\pi\times 6835~\rm MHz$ and  $\tau=100~\mu\rm s$ gives $\omega_{ba}\tau\sim 4 \times 10^6$ which is sufficiently large for high fidelity operation.

There is an intrinsic source of error in this gate 
that scales with the ratio  $\Delta_{dd}/\Omega_R.$ Assume the Raman fields  used for the $|b\rangle\rightarrow|r\rangle$ transfer provide a $\pi$ pulse when at most one of the atoms is excited to $|r\rangle.$ Then if both atoms start in state $|b\rangle$ the excitation to $|r\rangle$ will be imperfect due to the dipole-dipole shift. In this case the gate operation will end with a small amplitude for the atoms to remain in state $|rr\rangle.$   It is readily shown that the probability
for this to happen, and hence the gate error, is $E\sim    (\Delta_{dd}/\Omega_R)^2.$ This error is intrinsic to the design of the gate, so we require a ratio of interaction shifts to Rabi frequency of $0.1$, or less, for high fidelity operation.  

Unfortunately, the finite lifetime of the Rydberg state gives a finite decoherence rate, and an error that grows linearly with the time of the gate.  As the gate time scales with $1/\Delta_{dd}$ the interaction must be sufficiently strong to achieve high fidelity. We can write the probability of decoherence as $P\sim  t_{dd}/\tau =  \pi/\tau \Delta_{dd}$ where $\tau$ is the excited state lifetime 
 due to all decay mechanisms (see Secs. \ref{sec.qubit2radiative}, \ref{sec.qubit2phi}). The total gate error accounting for both imperfect fidelity and decoherence is then $E\sim  (\Delta_{dd}/\Omega_R)^2 + 1 /(\tau\Delta_{dd}).$ The error is minimized for $\Delta_{dd}\sim( \Omega_R^2/\tau)^{1/3}$, which then gives $E \sim (1/\tau\Omega_R)^{2/3}.$ We see that excited state lifetime  dictates how  fast the
Rydberg excitation must be performed to achieve a desired error level. 

\begin{table}[t]
\setlength{\extrarowheight}{5pt}
\centering
\begin{tabular}{|l|c|c|}
\hline
input state & decoherence error & rotation error\\
\toprule
$|aa\rangle$ & $\frac{2\pi|\Omega_R|}{\tau\omega_{ba}^2}( 1+\frac{|\Omega_R|}{2\Delta_{dd}} )$  &$\frac{2|\Omega_R|^2}{\omega_{ba}^2}$\\

$|ab\rangle$ or  $|ba\rangle$ &$\frac{\pi|\Omega_R|}{\tau\omega_{ba}^2}(1+\frac{|\Omega_R|}{2\Delta_{dd}} )+\frac{\pi}{\tau\Delta_{dd}}(1+\frac{\Delta_{dd}}{|\Omega_R|} 
)$&$\frac{|\Omega_R|^2}{\omega_{ba}^2}$ \\

$|bb\rangle$ & $\frac{2\pi}{\tau\Delta_{dd}}(1+\frac{\Delta_{dd}}{|\Omega_R|} 
)$& $\frac{8\Delta_{dd}^2}{|\Omega_R|^2}$\\
\hline
average &$\frac{\pi|\Omega_R|}{\tau\omega_{ba}^2}(1+\frac{|\Omega_R|}{4\Delta_{dd}})+\frac{\pi}{\tau\Delta_{dd}}(1+\frac{\Delta_{dd}}{|\Omega_R|} 
)$ & $\frac{|\Omega_R|^2}{\omega_{ba}^2}+\frac{2\Delta_{dd}^2}{|\Omega_R|^2}$\\
\hline
\end{tabular}
\caption{Leading conributions to the fidelity errors of a phase gate in the large Rabi frequency limit. See text for details. }
\label{tab.errorlargerabi}
\end{table}

\subsection{Large Rabi frequency}
\label{sec.qubit2rabilarge}

In this subsection we investigate operation in the large Rabi frequency limit in detail. 
To quantify the fidelity error we average over the four possible initial states of the phase gate which gives the results shown in Table \ref{tab.errorlargerabi}.
The calculations are performed using the rotation matrix of Eq. (\ref{eq.Rlong}) and the fidelity definition of Eq. (\ref{eq.fidelity}).  The decoherence error listed in the table is the integrated probability of a transition out of the Rydberg state during the gate operation due to spontaneous emission or other mechanisms.  For example the decoherence error for each atom
with the initial condition $|aa\rangle$ is $E\simeq(2/\tau)\int_0^{\pi/|\Omega_R|}dt~(|\Omega_R|^2/\Omega'^2)\sin^2(\Omega' t/2)+\pi/(\Delta_{dd}2\tau),$ with $\Omega'=\sqrt{|\Omega_R|^2+\omega_{ba}^2}.$ The error is then doubled to account for two atoms.  

The rotation 
error corresponds to the probability that the atom is in the Rydberg state at the end of the gate. For the initial condition $|bb\rangle$ the two atom error is 
\begin{eqnarray}
E=1-\left|\begin{pmatrix}1 \\ 0 \end{pmatrix}^T R(\Omega_R,\Delta_{dd},\pi/|\Omega_R|)\begin{pmatrix}1 & 0\\ 0 & -1 \end{pmatrix} \right.\nonumber \\
\left. \times ~~
R(\Omega_R,\Delta_{dd},\pi/|\Omega_R|)
\begin{pmatrix}1 \\ 0 \end{pmatrix}  \right|^4.
\end{eqnarray}
Here the rotation matrices operate on the states $|b\rangle$ and $|r\rangle$ and the  vector $\begin{pmatrix}1 & 0 \end{pmatrix}^T$ represents $|b\rangle.$  
The last row in Table \ref{tab.errorlargerabi} gives the gate error averaged over the possible input states. While a particular computation may weight certain input states more heavily than others the average gate error is indicative of the gate performance. 
The average error is minimized for 
$$
\Delta_{dd}|_{\rm opt} \simeq \left( \frac{\pi |\Omega_R|^2}{4\tau}\right)^{1/3}\left(1+\frac{|\Omega_R|^2}{4\omega_{ba}^2} \right)^{1/3} 
$$
which determines the leading order contributions to the error as 
\begin{equation}
E|_{\rm opt} \simeq \frac{|\Omega_R|^2}{\omega_{ba}^2}+ \frac{3}{2^{1/3}}\left( \frac{\pi }{\tau|\Omega_R|}\right)^{2/3}. 
\label{eq.eopt}
\end{equation}

\begin{figure}[!t]
\centering
\includegraphics[width=7.5cm]{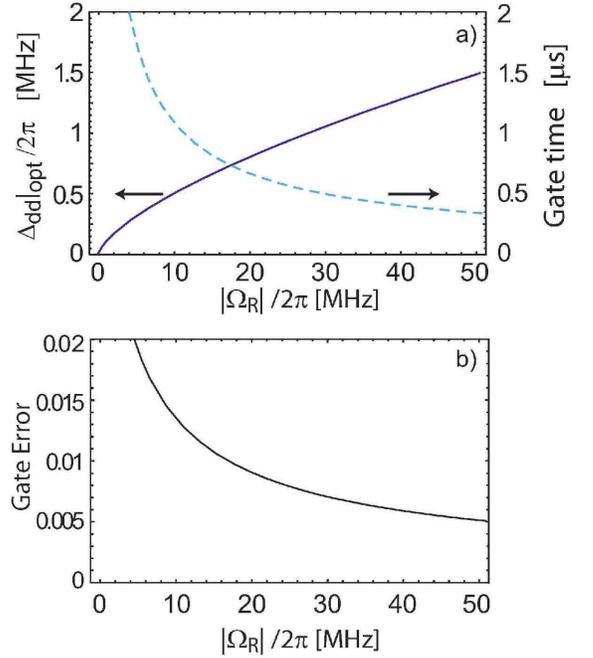}
\caption{(color online) Two-qubit phase gate performance in the limit of large Rabi frequency: a) optimum dipole-dipole shift and gate time, and b) minimum achievable fidelity error. Calculations for $\tau=100~\mu\rm s$ and $\omega_{ba}=2\pi\times 6835~\rm MHz.$  }
\label{fig.largerabierror}
\end{figure}

The optimum dipole-dipole shift, gate time gate time $t_g=2\pi/|\Omega_R|+ \pi/\Delta_{dd}|_{\rm opt},$ and gate error are shown in Fig. \ref{fig.largerabierror}.
 We see that errors less than $10^{-2}$ are possible with Rabi frequencies of a few tens of MHz and a gate time of less than $1~\mu\rm s.$ The corresponding optimum dipole-dipole shifts of order $1 ~\rm MHz$ can be achieved at tens of microns of separation as we discuss in Sec. \ref{sec.dipoledipole}.

Finally we note that the absolute minimum gate error occurs for $|\Omega_R|$ intermediate between 
$\omega_{ba}$ and $1/\tau.$ Minimizing Eq. (\ref{eq.eopt}) with respect to $|\Omega_R|$ we find 
$|\Omega_R|_{\rm opt}=[\pi\omega_{ba}^3/(2^{1/2}\tau)]^{1/4}$ and a minimum error of 
$E|_{\rm min}=[2^{7/2}\pi/(\tau \omega_{ba})]^{1/2}.$ For the parameters used in Fig. \ref{fig.largerabierror} we get $|\Omega_R|_{\rm opt}=2\pi \times 183~\rm MHz$ and $E|_{\rm min}=2.9\times 10^{-3}.$

\subsubsection{Additional imperfections}
\label{sec.gateforceheating}

There are two additional imperfections when working in the large Rabi frequency limit.
The first of these is due to fluctuations in the distance between the atoms.   For atomic separation large compared to the extent of the thermal motion  the actual qubit separation will be $R+x_1+x_2$, with $\left< x_1^2\right> =\left< x_2^2\right> =(w_{f0}^2/4)(T_a/|U_m|).$
The variance of the interaction phase is easily shown to be $\Delta^2(\phi)=(9\pi^2/2)(w_{f0}^2/R^2)(T_a/|U_m|). $ Using this variance as a measure of the gate error gives 
$E=  (1/4)\Delta^2(\phi)\times(1/4)=3.5\times 10^{-4}$ at $R=50~\mu\rm m.$ The final factor of $1/4$ accounts for an averaging over the possible initial states. We see that at $R=50~\mu\rm m$ this error is small compared to the intrinsic gate error which dominates for $R\gtrsim 15~\mu\rm m.$ 

The second subsidiary imperfection is the presence of heating due to two-body forces when both atoms are excited to the Rydberg state.  This effect could limit  the number of gate operations before motional cooling is needed and lead to decoherence through undesired entanglement of the motional and spin states. The heating rate can be estimated simply as $P=Fv\sim F\omega_f w_{f0}.$ Using $F=-dV_{dd}/dR$ for the two body force we get a peak heating power of 
$P\sim 3\hbar \Delta_{dd}\omega_x w_{f0}/R.$ Using $\Delta_{dd}=2\pi\times 1 \rm~ MHz$ and $R=50~\mu\rm m$ gives $P\sim  1.8 ~\mu{\rm K}/\mu \rm s $ for our standard FORT parameters.  We therefore expect a maximum of about 1 $\mu K$ of heating when both atoms are initially in the state $|b\rangle$ which is coupled to the Rydberg states. Although the heating power will average to zero over many operations there is a finite probability for undesired motional entanglement.  The  spacing of the radial vibrational levels in temperature units is $\Delta E_{\rm vib}=\hbar\omega_x/k_B=1.9 ~\mu\rm K$ which is comparable to the peak heating value. On average there will be a reduced probability for a change of the vibrational state since the atoms spend proportionately more time near the turning points of the motion where the velocity is small.  

The above errors due to fluctuations in the atomic separation and two-body forces are specific to gate operation in the large Rabi frequency limit. There is an additional error source that is common to both modes of gate operation, which is the presence of Doppler shifts of the Rydberg excitation beams  due to atomic motion. For ground state qubit rotations two-photon stimulated Raman transitions are essentially Doppler free for co-propagating beams, and the error given by Eq. (\ref{eq.dopplererror}) is insignificant. Two-photon excitation of Rydberg levels using $0.78~\mu\rm m$ and $0.48~\mu\rm m$ beams as indicated in Fig. \ref{fig.doubleqgeometry} cannot be made Doppler free. For $T_a=50~\mu\rm K$ and $|\Omega_R|=2\pi\times 10 ~\rm MHz$ we find using   Eq. (\ref{eq.dopplererror}), fidelity errors of $3.3 \times 10^{-4}$ and $1.9\times 10^{-5}$ for co- and counter-propagating beams respectively. These errors are significantly smaller than the intrinsic gate errors shown in Figs. \ref{fig.largerabierror},\ref{fig.largedderror}.

\subsection{Large dipole-dipole frequency shift}
\label{sec.qubit2ddlarge}

There is an alternative mode of operation of the phase gate that removes the dependence on 
variations in the interatomic separation and eliminates the two body heating discussed above. This mode can be used for closely spaced atoms for which 
$\Delta_{dd}\gg |\Omega_R|.$ In this limit the  time scales  have the ordering 
$1/\omega_{ba}\ll  1/\Delta_{dd}\ll 1/|\Omega_R|\ll \tau$. The protocol for a conditional phase gate is then 
\cite{ref.cote}: 1) excite atom 1(control atom) from $|b\rangle\rightarrow |r\rangle$ with a $\pi$ pulse, 2) excite atom 2(target atom) from $|b\rangle\rightarrow |r\rangle\rightarrow |b\rangle$ with a $2\pi$ pulse,  and 3) transfer atom 1  back down from $|r\rangle\rightarrow |b\rangle$ with a $\pi$ pulse. Note that in contrast to the protocol used in the large Rabi frqeuency limit we assume here that the atoms can be individually addressed.


\begin{table}[t]
\centering
\begin{tabular}{|l|c|c|}
\hline
input state & decoherence error & rotation error\\
\toprule
$|aa\rangle$ & $\frac{2\pi|\Omega_R|}{\tau\omega_{ba}^2}$&$\frac{2|\Omega_R|^2}{\omega_{ba}^2}$\\

$|ab\rangle$  &$\frac{\pi}{\tau|\Omega_R|}\left(1+\frac{|\Omega_R|^2}{\omega_{ba}^2} \right)$&$\frac{|\Omega_R|^2}{\omega_{ba}^2}$ \\

$|ba\rangle$ &$\frac{\pi}{\tau\Delta_{dd}}+\frac{\pi}{\tau|\Omega_{R}|}\left(1+\frac{|\Omega_R|^2}{\omega_{ba}^2} 
\right)$&$\frac{|\Omega_R|^2}{2\omega_{ba}^2}$ \\

$|bb\rangle$ & $\frac{\pi}{\tau}\left(\frac{2}{|\Omega_R|}+\frac{1}{\Delta_{dd}} \right)$& $\frac{|\Omega_R|^2}{2\Delta_{dd}^2}$\\
\hline
average &$\frac{\pi}{\tau|\Omega_R|}\left(1+\frac{|\Omega_R|^2}{\omega_{ba}^2} \right)+\frac{\pi}{2\tau\Delta_{dd}}$ & $\frac{|\Omega_R|^2}{8\Delta_{dd}^2}+\frac{7|\Omega_{R}|^2}{8|\omega_{ba}|^2}$\\
\hline
\end{tabular}
\caption{Leading conributions to the fidelity errors of a phase gate in the large dipole-dipole shift limit. See text for details. }
\label{tab.errorlargedd}
\end{table}

The resulting idealized logic table is: $|aa\rangle\rightarrow|aa\rangle,$ 
$|ab\rangle\rightarrow |ab\rangle,$
$|ba\rangle\rightarrow |ba\rangle,$
$|bb\rangle\rightarrow -|bb\rangle$ which is an  entangling phase gate.  The gate errors are calculated using the same techniques as in Sec. \ref{sec.qubit2rabilarge} which results in the errors shown in Table \ref{tab.errorlargedd}. The average error is minimized for 
$$
\Omega_R|_{\rm opt} \simeq \left( \frac{4\pi \Delta_{dd}^2}{\tau}\right)^{1/3}  - 
\frac{4\pi\Delta_{dd}^2}{3\tau\omega_{ba}^2} 
$$
which determines the leading order contributions to the error as 
\begin{equation}
E|_{\rm opt} \simeq \frac{3 \pi^{2/3}}{2^{5/3}} \frac{1}{(\Delta_{dd}\tau)^{2/3}} 
\left(1+\frac{7\Delta_{dd}^2}{3\omega_{ba}^2}\right). 
\label{eq.eoptdd}
\end{equation}

\begin{figure}[!t]
\centering
\includegraphics[width=7.5cm]{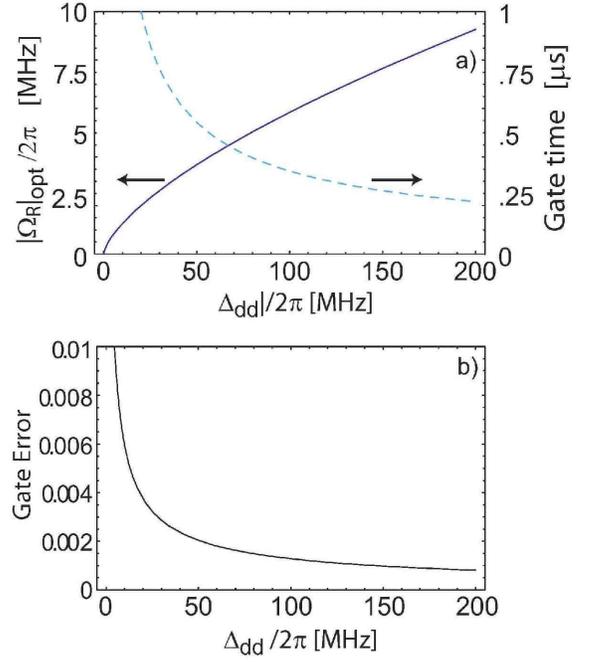}
\caption{(color online) Two-qubit phase gate performance in the limit of large dipole-dipole shift: a) optimum Rabi frequency and gate time, and b) minimum achievable fidelity error. Calculations for $\tau=100~\mu\rm s$ and $\omega_{ba}=2\pi\times 6835~\rm MHz.$  }
\label{fig.largedderror}
\end{figure}

The optimum Rabi frequency, gate time $t_g=4\pi/|\Omega_R|_{\rm opt}|,$ and gate error are shown in Fig. \ref{fig.largedderror}.
 We see that errors much less than $10^{-2}$ are possible with dipole-dipole shifts  of a few tens of MHz and a gate time of less than $1~\mu\rm s.$ This protocol appears more promising for achieving very small gate errors than the  large Rabi frequency protocol, since it is easier to achieve  very large dipole-dipole shifts than it is to achieve very large Rabi frequencies. 

Finally we note that minimizing Eq. (\ref{eq.eoptdd}) with respect to $\Delta_{dd}$ we find 
$\Delta_{dd}|_{\rm opt}=\sqrt{3/14}\omega_{ba}$ and an absolute  minimum error of 
$E|_{\rm min}=[1701\pi^2/(128 \tau^2 \omega_{ba}^2)]^{1/3}.$ For the parameters used in Fig. \ref{fig.largedderror} we get $\Delta_{dd}|_{\rm opt}=2\pi \times 3160~\rm MHz$ and $E|_{\rm min}=1.9\times 10^{-4}.$

\subsection{Dipole-dipole interaction strength}
\label{sec.dipoledipole}

The long-range  interactions between two Rydberg atoms are extremely
strong and are the heart of the quantum computation scheme discussed in
this paper.  We consider here the Rydberg-Rydberg interactions in two
limits: zero electric field, where the long-range $1/R^6$ van-der-Waals
interactions are of primary importance, and in a hybridizing electric
field, where the atoms attain a ``permanent'' dipole moment of magnitude
$n^2ea_0$.

\begin{figure}[!b]
\centering\includegraphics[width=8.cm]{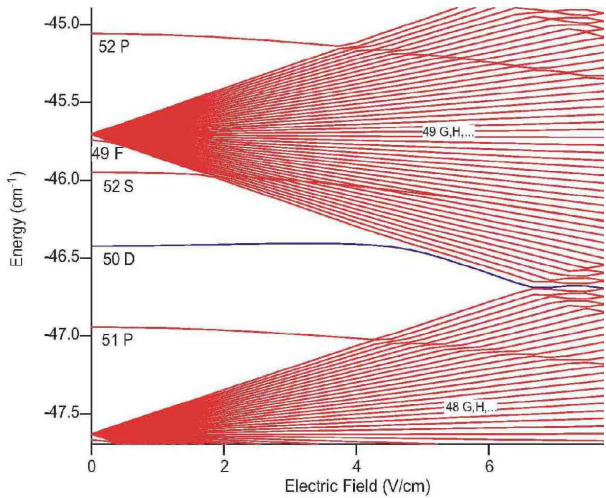}
\caption{(color online) Stark map for states near $n=50$ for $^{87}$Rb.  At electric fields 
around 5-6 V/cm, the atom acquires a large permanent
dipole moment, oriented in space along the applied field. }\label{starkmap}
\end{figure}

\begin{figure}[!t]
\centering\includegraphics[width=8.cm]{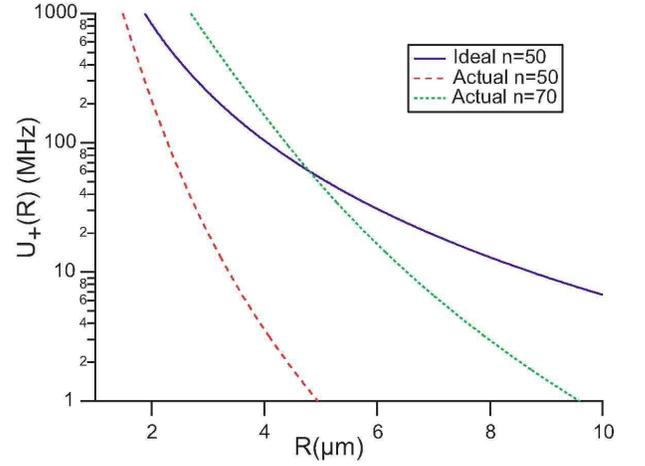}
\caption{(color online) Isotropic dipole-dipole interaction for excitation to the Rb 
50S state.  The energy defect between the 50S+50S and
49P+50P states significantly reduces the interaction as compared to 
the ideal degenerate case. Going to larger $n$
partially compensates for the non-zero energy defect.}\label{pot50s}
\end{figure}

For S states, the van-der-Waals interaction is dominated by the near
resonance between the states $nS+nS$ and $nP+(n-1)P$, as can be seen
from Fig.~\ref{starkmap}. In the approximation that we neglect the
contribution from other states, the Rydberg-Rydberg potential energy
curves are simply given by
\begin{equation}
V(R)=\frac{\delta }{2} + {\sqrt{\frac{4U_3(R)^2}{3} + \frac{{\delta
}^2}{4}}}
\end{equation}
where 
$U_3(R)=e^2\langle50S||r||50P\rangle\langle50S||r||49P\rangle/(4\pi\epsilon_0 R^3)=5.75\times10^3$ 
MHz $\mu$m$^3/R^3$ and the SS-PP energy
defect is $\delta=E(49P)+E(50P)-2E(50S)=-3000$ MHz.  At the 10 $\mu$m
separations of interest here, $U_3\ll\delta$ so that
\begin{equation}
V(R)=-\frac{4U_3(R)^2}{ 3\delta}
\end{equation}
As can be seen from Fig.~\ref{pot50s}, the van-der-Waals potential for
$50S$ is insufficient to allow for the MHz processing that is desired
here.

 Van-der-Waals interactions between D-states are dominated by the
near-resonance of $nD+nD$ and $(n+1)P+(n-1)F$.  In general, the
van-der-Waals interactions are of comparable strength when compared to
the S-states. Because of the degeneracy of the
$D$ levels, however, it can be shown that $U_3$ vanishes for one each of
the
$^3\Sigma_u^+$ and $^1\Sigma_g^+$ molecular states\cite{Walker04}.  Thus
there is no blockade at zero field for these states.

The scaling of the van-der-Waals interaction with $n$ is very rapid,
$V\propto n^{11}$.  For $n=100$, $V(10$ $\mu$m$)\sim50$ MHz. While this is 
sufficient for a conditional phase gate (see Fig. \ref{fig.largedderror}) we can  considerably 
 enhance the interaction strength by applying a dc field.  Thus we now
consider the long-range interactions in the presence of an applied
electric field.

There are two primary ways to enhance the Rydberg-Rydberg interaction
using an applied electric field.  The first is to use the electric 
field dependence of the state energies to bring the energy defect
$\delta$ to zero, in which case
$V(R)\approx{2U_3(R)/\sqrt{3}}
$.  This would produce an extremely strong isotropic interaction of 
maximum strength, for example $n=95$ gives a frequency
shift at 10
$\mu$m separation of 160 MHz.  However, inspection of 
Fig.~\ref{starkmap} shows that the P-states tune the wrong way in an 
electric
field, increasing rather than decreasing $\delta$.  For D-states 
there remains the problem of zero dipole-dipole interactions for some
molecular symmetries.

The second way to adjust the Rydberg-Rydberg interactions is to use a 
large enough field to strongly mix states of different $L$.  In
this case the atom acquires a field-independent dipole moment, as 
suggested for example  by the linear field dependence of the 50$D$
state at electric fields between 5 and 6 V/cm.  In this case the 
Rydberg-Rydberg interaction is strong but anisotropic and we have the dipole-dipole 
interaction of Eq. (\ref{eq.deltadd})
\begin{eqnarray}
V_{dd}(R)=\hbar \Delta_{dd}(R)
&=&\frac{\mu^2}{4\pi\epsilon_0 R^3}(1-3\cos^2\theta)
\end{eqnarray}
where $\theta$ is the angle between the interatomic axis and the 
electric field.

\begin{figure}[!t]
\centering
\includegraphics[width=8.cm]{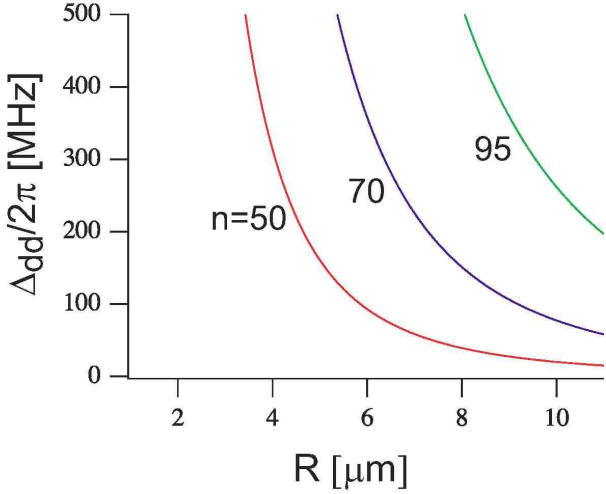}
\caption{(color online) Dipole-dipole interaction strengths as a function of 
interatomic distance for various principal quantum
numbers. }
\label{fieldpots}
\end{figure}

In the proposed geometry for the quantum computer, the electric 
fields can be aligned to $\theta=0$ for maximum interaction strength.
The electric dipole moment is on the order of $n^2 e a_0$, giving a 
dipole-dipole interaction of comparable size as the zero field
case with $\delta=0$.  We have numerically estimated the dipole 
moments for the field-mixed D-states, and obtain for example $\mu=3300 ~ea_0$ for $n=50.$ 
  The resulting interaction strengths as a function 
of distance are shown in Fig.~\ref{fieldpots}. We see that 
interaction frequencies in excess of 100 MHz can be achieved for $R=8~\mu\rm m$ at $n=70$ 
and well beyond  $R=10~\mu\rm m$ at $n=95.$ 
  Thus the dipole-dipole interaction strength needed to optimize the phase gate, as shown in  Fig. \ref{fig.largedderror},
can be achieved for qubit separations tht are optically resolvable. It has been pointed out\cite{ref.ryabtsev} that $n$ should not be too large in order to avoid collisions between the valence electrons of neighboring Rydberg atoms. For $R=8~\mu\rm m$ this implies $n$ should be less than about 100. Although $n=100$ is sufficient for the geometry considered here, it is also true that in the large dipole-dipole shift limit discussed above this limitation is not needed as only one atom at a time is excited to a Rydberg level.

\subsection{Balancing the ground and excited state polarizabilities}
\label{sec.balancepolarize}

\begin{figure}[!t]
\centering
\includegraphics[width=8.8cm]{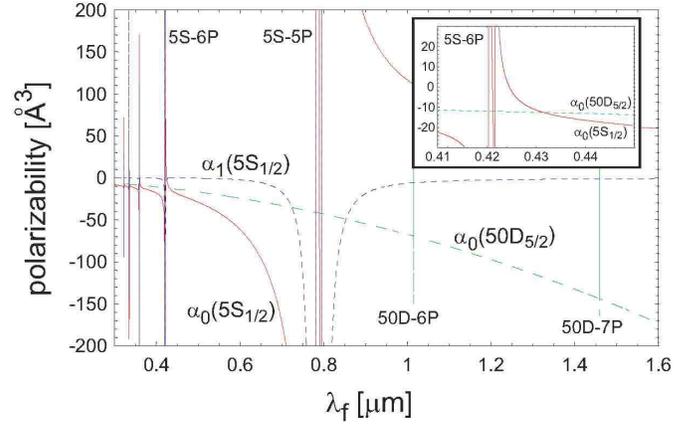}
\caption{(color online)  Polarizability of the $5S_{1/2}$ and $50D_{5/2}$ states. The inset shows the behavior near the $5S - 6P$ transition.  }
\label{fig.polarizability}
\end{figure}

The implementation of the two qubit conditional gate described so far has an inherent flaw that stems from the 
need to make real transitions to Rydberg levels. The atoms are confined in attractive optical potential wells since the ground state polarizability is positive for light tuned to the red of the first Rb resonance lines.  However, for the same red detuned trapping laser, the  high lying Rydberg levels have a negative polarizability  that provides a repulsive potential. Excitation of an atom to the repulsive Rydberg   state  during a  gate cycle leads to heating and decoherence through entanglement of the spin and motional states.  Looking at Fig. \ref{fig.polarizability} we see that for $\lambda_f=1.06~\mu\rm m$ the $50D_{5/2}$ Rydberg level has a polarizability of $\alpha_0(50D_{5/2})=-77~ \AA^3$. This polarizability is about 95\% of the free electron polarizability of $-e^2/m_e\omega_f^2$, with $m_e$ the electron mass. This implies that the Rydberg polarizability is only weakly dependent on the principal quantum number $n$ for large $n$, so the discussion given here for the case of the $50D_{5/2}$ level is applicable to any of the highly excited states.   

When the Rydberg state sees a repulsive potential the heating is minimized by turning off the trapping laser during the gate operation. For atomic temperatures large compared to the trap vibrational energy the motion is quasi-classical and the average heating per gate cycle of duration 
$\tau$ is $\left<\Delta U\right>=T_a \frac{1}{2}(\omega\tau)^2,$ where $\omega$ is the radial trap vibration frequency. For $\tau=0.5 ~\mu\rm s$ and $\omega = 2\pi \times 39~\rm kHz$ 
 we get $\left<\Delta U\right>/T_a=7.5\times 10^{-3}.$ This heating rate is about 5 times  larger than that due to the AC Stark shifts of the Raman lasers discussed in Sec. \ref{sec.acstarkshifts}, and implies that several hundred operations can be performed before there is significant heating of the atomic motion. 
 
The heating can be eliminated by choosing a FORT laser wavelength that
gives equal polarizability for the ground and Rydberg states. Referring to Fig. 
\ref{fig.polarizability}, one possibility  is to tune the FORT laser to the blue side of the $50D-6P$ transition where the Rydberg level acquires a positive polarizability.   
The scalar polarizability of the Rydberg level is given by
\begin{equation}
\alpha_0(nD_{5/2})= -\frac{2}{3\hbar}\frac{1}{2 J +1} \sum_{\gamma' J'} \frac{\omega_{\gamma' J',\gamma J}}{\omega_{\gamma' J',\gamma J}^2 -\omega_f^2}|\langle \gamma' J' ||\hat  D ||\gamma J\rangle|^2,
\label{eq.alpharydberg}
\end{equation}
where $\hat D$ is the electric dipole operator and $\hbar \omega_{\gamma' J',\gamma J}=E_{\gamma' J'}-E_{\gamma J}.$ 
Away from the resonance the polarizability changes slowly with $\omega_f$ and for the $|50D_{5/2}\rangle$
level it is $\bar\alpha_0(50D_{3/2})\simeq -69~\AA^3$.
When $\omega_f$ is near resonant with the $|n D_{5/2}\rangle \leftrightarrow |6P_{3/2}\rangle$ transition 
the polarizability is the sum of the background value $\bar\alpha_0$ plus the resonant contribution to the sum.   Using Coulomb wavefunctions  for $n=50$ we find $\langle 50D_{5/2} || \hat D ||6P_{3/2}\rangle= 0.059 ~e a_0 $ 
so that a rough estimate for the detuning condition at which the polarizabilities are equal is 
\begin{eqnarray}
\Delta_f&=&\frac{1}{3\hbar}\frac{1}{2J+1}\frac{|\langle 50D_{5/2} ||\hat  D ||6P_{3/2}\rangle|^2}{\alpha_0(5S_{1/2})-\bar\alpha_0(50D_{5/2})}\nonumber\\
&=&  2\pi \times 945  ~\rm MHz.
\end{eqnarray}

Unfortunately there is a penalty associated with balancing the polarizabilities in this way  since there is a  probability for the gate operation to end with the atom having finite amplitude to be in the   $|6P_{3/2}\rangle$ level which lies outside the computational basis. 
The Rabi frequency for this transition (starting from $m_J=3/2$) is given by $|\Omega_f|=({\mathcal E}_f/(\hbar\sqrt 15))\langle 50D_{5/2} ||\hat  D ||6P_{3/2}\rangle $ and the probability for population transfer to the $|6P_{3/2}\rangle$ state at the end of a  Rydberg operation is bounded by $P_{\rm max}=|\Omega_f|^2/(|\Omega_f|^2 + \Delta_f^2).$ For a trap depth of 1 mK we find $|\Omega_f|= 2\pi\times 210~\rm MHz$ and 
$P_{\rm max}=0.16.$ This upper bound assumes that the product of the gate time and the spontaneous lifetime of the $6P$ level is not large, otherwise there is an additional loss mechanism  due to decay out of the 
$6P$ level. 

This decoherence probability is roughly proportional to the FORT laser intensity and can be reduced by working with a shallower FORT trap. This suggests that atoms be loaded into 1 mK deep traps where they can be cooled to sub Doppler temperatures  with standard methods, followed by adiabatic reduction of the trap depth before performing logical operations. With $T_a=5~\mu\rm K$ and $|U_m|=100 ~\mu\rm K$ the atom localization and motional errors will be the same as considered in the rest of this paper, while the decoherence probability per Rydberg operation will be bounded by  $P_{\rm max}=0.016.$ 

While  we expect this to be a viable approach for initial experiments it can be readily shown that the decoherence probability with polarizability balancing of  a $nD$ Rydberg level scales proportional to $1/|\langle nD_{5/2} ||\hat  D ||6P_{3/2}\rangle|^2$.  Since $|\langle nD_{5/2} ||\hat  D ||6P_{3/2}\rangle|^2\sim 1/n^3$ the leakage problem scales as $n^3$ and the  technique is most useful for low lying Rydberg states that have large transition dipole moments with the $6P$ level. 

An alternative solution is to choose a FORT wavelength such that the ground state acquires a negative polarizability which coincides with that of the Rydberg level.
In this situation the optical potential is negative so  the atom is trapped at a local minimum of the intensity. This approach has a number of advantages since for an intensity profile of the form $I=I_0(1-e^{-2r^2/w_{f0}^2})$  the time averaged trapping intensity at the position  of the atom is $T_a/|U_m|$ times smaller than it would be for an attractive potential of the same depth.  This increases the $T1$ and $T2$ times in rows 2,3 and 6 of Table \ref{tab.groundstate} associated with photon scattering and motional AC Stark shifts by the same factor.  
 It was pointed out by Safronova et al. \cite{ref.safronova} that the polarizabilities can be balanced in this way by tuning the trapping laser between the D lines. For $^{87}$Rb the balancing point is at $\lambda_f\simeq.79~\mu\rm m$. 
 However the vector polarizability at this point has a very large value of
$\alpha_1\simeq 3600~\AA^3$ which would lead to 
a $T1$ due to inelastic scattering (see Eq. \ref{eq.T1raman}) of about $1~\rm ms$ for the parameters of Table \ref{tab.groundstate},  which is unacceptably short.  

A more favorable alternative is to use a FORT wavelength of $\lambda_f\simeq0.431~\mu\rm m$ which gives equal polarizabilities of $\alpha_0\simeq -12.5~\AA^3,$
as shown in the inset to Fig. \ref{fig.polarizability}. At this wavelength 
$\alpha_1\simeq -0.01~\AA^3$ which gives an inelastic scattering $T1$ that is longer than any other time scale in Table \ref{tab.groundstate}.  The only drawbacks of this solution are technical, not fundamental. There is  additional experimental complexity associated with creating optical beams with local intensity minima, as well as the need for a medium power laser in the blue. Recent developments in solid state laser sources and parametric frequency convertors render the latter requirement readily solvable.


\subsection{Rydberg state radiative lifetimes}
\label{sec.qubit2radiative}

The lifetimes of  highly excited Rydberg $nl$ states are affected strongly by background blackbody radiation. 
If the 0 K lifetime is $\tau_{nl}^{(0)}$ the finite temperature lifetime can be written as 
\begin{equation}
\frac{1}{\tau_{nL}}=\frac{1}{\tau_{nL}^{(0)}}+\frac{1}{\tau_{nL}^{(\rm bb)}}
\end{equation}
where $\tau_{nL}^{(\rm bb)}$ is the finite temperature blackbody contribution. 
The 0 K radiative lifetime can be calculated by summing over transition rates or approximated by the expression\cite{ref.gounand} 
\begin{equation}
\tau_{nL}^{(0)}=\tau_L^{(0)} (n^*)^{\alpha_L}.
\end{equation}
For all the  alkalis $\alpha_L\simeq 3.$ Parameters for Rb are 
$\alpha_S=2.94,$ $\alpha_P=3.02,$ $\alpha_D=2.85,$ $\alpha_F=2.95,$  
and $\tau_S^{(0)}=1.43$, $\tau_P^{(0)}=2.76$, $\tau_D^{(0)}=2.09$, and $\tau_F^{(0)}=.76$ in units of ns. 

\begin{figure}[!t]
\begin{minipage}[c]{8.cm}
\includegraphics[width=8.cm]{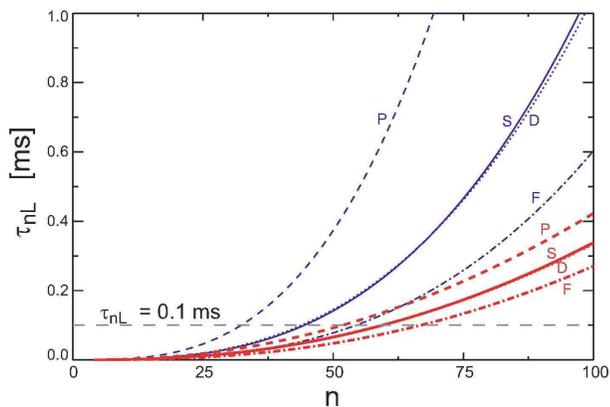}
\caption{(color online) Excited state lifetime due to radiative decay  for $T=0$ (blue, thin lines) and $T=300$ K (red, thick lines) for S,P,D, and F states of Rb.  }
\label{fig.raddecay}
\end{minipage}
\end{figure}

For large $n$ the blackbody rate can be written approximately as\cite{ref.gallagherbook}
\begin{equation}
\frac{1}{\tau_{nL}^{(\rm bb)}}=\frac{4\alpha^3 k_BT}{3\hbar n^2},
\label{eq.bbrate}
\end{equation} 
where $\alpha$ is the fine structure constant. 
Equation (\ref{eq.bbrate}) includes transitions to continuum states so that it accounts for blackbody induced photoionization. 
Figure \ref{fig.raddecay} shows the radiative lifetime   $\tau_{nL}$ for $n$ up to 100 and several $l$ states. 
We see that for $n\gtrsim65$  the S,P,D, and F states have lifetimes greater than $0.1~\rm ms$ at room temperature.

\subsection{FORT trap induced photoionization}

\label{sec.qubit2phi}

Highly excited states are also unstable against photoionization from the
intense trapping light.  Since the Rydberg
electron is nearly free, and the photoionization is far above threshold,
the photoionization cross sections are small.  We have estimated the
photoionization rate of high Rydberg states by calculating the cross
section using Rydberg and continuum wavefunctions.

\begin{figure}[!tb]
\includegraphics[width=8.cm]{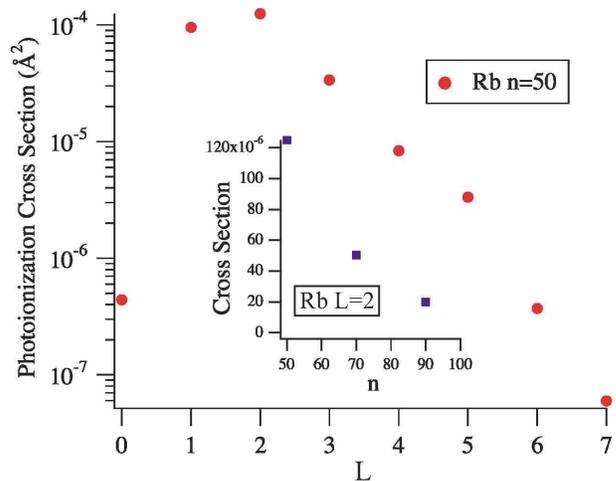}
\caption{(color online) Photoionization cross section vs ejected electron
energy, for the Rb 50S Rydberg state with $\lambda_f=1.01~\mu\rm m$.}\label{fig.photocross}
\end{figure}

The cross section is
given by  \cite{ref.gallagherbook}
\begin{equation}
\sigma={2\pi^2}\frac{\hbar e^2}{ mc}\left.\frac{df}{dE}\right|_{E_c}
\end{equation}
where $f$ is the oscillator strength and $E_c=\hbar\omega+E_r$ the energy
of the continuum electron produced by the photon of frequency $\omega$
from the Rydberg state of energy $E_r$.  The oscillator strength
distribution is
\begin{equation}
\frac{df}{dE}=
\sum_L  \frac{2m\omega L_>}{ 3\hbar (2L_r+1)}\left|\int\Psi_r(r) r
\phi_{L,E}(r) dr\right|^2
\end{equation}
where $L_>$ is the larger of $L_r$ and $L$ and the continuum wavefunction is normalized per unit energy
\begin{equation}
\phi_{L,E}(r)\stackrel{r\rightarrow\infty}{\longrightarrow} 
\sqrt{\frac{2m}{\pi \hbar^2 k}}\sin(kr+\delta_{L,E}).
\end{equation}
We have used two methods to detemine the bound and continuum wavefunctions. In the first we use quantum defect theory\cite{ref.seaton} to find the Coulomb wavefunctions and phase shifts $\delta_{L,E}$. The second method uses  the
L-dependent model potentials of Marinescu {\it et al.}\cite{Marinescu94b},
with values slightly adjusted to give the proper quantum defects of the
Rydberg states.  The Numerov method was then used to find $\phi_{L,E}$. Numerical results for the two approaches agree within a factor of about 2 for the range of $n$ discussed here.  The discussion and Fig. \ref{fig.photocross} give the results obtained using the Marinescu potential.

The $n=50$ photoionization cross sections are shown as a function of $L$
in Figure~\ref{fig.photocross}.  The photoionization cross section for the
S-state is much smaller than $L>1$ states due to the $\pi$/2 phase
shift between the S and P wavefunctions \cite{ref.gallagherbook}.  The cross
sections increase dramatically for the $P$ and $D$ states before slowly
decreasing with further increases in $L$.  While the 50S cross sections
are very small, the 100$\times$ larger cross section for the higher $l$
levels implies that the photoionization rate for the $S-$states will depend
sensitively on mixing with the $P-$levels due to external fields.  Thus the
photoionization cross sections may be as large as 10$^{-20}$ cm$^2$, or
as high as $\gamma_{\rm pi}=$ 31,000/s for a 1 mK trap depth.

The photoionization cross sections decrease with $n$, a factor of
6 going from $n=50$ to $n=90$ as shown in the inset to
Figure~\ref{fig.photocross}. Thus at $n=90$ we get a minimum lifetime of $\tau_{\rm pi}\sim  190 ~\mu\rm s$
for the $P$ and $D$ states and substantially longer for the other $L$ states. 
Taking into account the blackbody lifetime calculated in the previous section  we conclude that the  Rydberg lifetime will exceed $100 ~\mu\rm s$ for $n\gtrsim 80.$ This validates the gate fidelity estimates discussed in Secs. 
\ref{sec.qubit2rabilarge} and \ref{sec.qubit2ddlarge}. Since the photoionization rate scales linearly with the trap depth even better performance is possible by further cooling of the atomic motion and a corresponding reduction of the trap depth, or by use of a blue detuned FORT laser as discussed in Sec. \ref{sec.balancepolarize}.

\section{Discussion}
\label{sec.discussion}In this paper we have presented a detailed analysis of  quantum logic 
using neutral atoms localized in optical traps. Two-photon Raman transitions are used for single qubit gates, dipole-dipole interactions of Rydberg states provide a two-qubit conditional phase gate, and detection of resonance fluorescence is used for state measurement. 
We have concentrated on an implementation of neutral atom gates that should be feasible with currently available experimental methods and laser sources. 
The $T1$ and $T2$ coherence times for qubit storage, and error estimates for single qubit gates are summarized in Tables \ref{tab.groundstate} and \ref{tab.singlequbit}. The intrinsic two-qubit gate errors are shown in Figs. \ref{fig.largerabierror} and \ref{fig.largedderror}.
The analysis supports 
the feasibility of MHz rate logical operations with intrinsic errors  $O(10^{-4})$ for single qubit operations, $O(10^{-3})$ for two qubit gates, and state measurements in less than $100 ~\mu\rm s$ with $O(10^{-3})$ measurement error.  We show that coherence times of at least several seconds are possible in red-detuned attractive optical traps.
  Taken together these numbers  
suggest  an attractive   framework for experimental studies of quantum logic with neutral atoms. 

It should be emphasized that the Rydberg gate approach does not rely on cooling the atoms to the motional ground state. While ground state cooling has been demonstrated in optical traps, and lower atomic temperatures will lengthen coherence times and reduce some of the gate errors, the difficulty of maintaining the atoms in the motional ground state in the presence of heating mechanisms should not be overlooked. Our assumption of $T_a=50~\mu\rm K$ does not require complex cooling schemes, and implies that $O(10^3)$ single qubit logical operations can be performed without significant reduction in fidelity due to motional heating.

 As discussed in 
Sec. \ref{sec.balancepolarize} the differential polarizability of the ground and Rydberg states in a red detuned FORT will lead to substantial heating or loss of coherence. While initial experiments 
are viable in red detuned FORTs our analysis suggests that a blue FORT where the atoms are trapped at a local minimum of the intensity will ultimately be necessary to realize the full potential of this scheme. The blue FORT will also substantially improve the   coherence times for qubit storage, and to a lesser extent the Rydberg state lifetime.

Extending the two-qubit approach described here to a large number of qubits will involve solving challenges related to loading and addressing of multiple sites. The Rydberg gate approach does appear intrinsically well suited for implementation in a two-dimensional array, including error correction blocks. The large dipole-dipole shift limit can be used for gates between neighboring sites, while the large Rabi frequency limit which works at longer range may allow non-nearest neighbors to be coupled. Figure \ref{fig.largerabierror} shows that in the limit of large Rabi frqeuency, gate errors less than $10^{-2}$ are possible with dipole-dipole coupling strengths of only 1 MHz. For Rydberg levels with $n=70$ we can achieve a coupling strength of 1 MHz at a separation of about $40~\mu\rm m.$ With a $8~\mu\rm m$ qubit spacing this suggests the possibility of coupling blocks of 25 or more qubits without physical motion. 
By taking advantage of the directional properties of the dipole-dipole interaction described by Eq. (\ref{eq.deltadd}) it is possible to perform row parallel operations between pairs of qubits with strongly suppressed crosstalk. While there are many appealing features of the approach studied here we emphasize that the extent to which it will  prove possible to perform arbitrarily large, scalable quantum computations, remains an open question that will require a great deal of further theoretical and experimental studies. 

\acknowledgments

We thank the members of the Rydberg atom quantum computing group at Madison  
for helpful discussions. 
This work is supported by the U. S. Army
Research Office under contract number DAAD19-02-1-
0083, and NSF grants EIA-0218341 and PHY-0205236.

\appendix 

\section{Raman transitions due to  imperfect polarization}
\label{sec.appendixpolarization}

We account for imperfect polarization of the Raman beams by putting $\bfepsilon_1= ({\bf e}_+ + \epsilon_{1-}{\bf e}_- + \epsilon_{10} {\bf e}_0)/(1+\epsilon_{1-}^2 + \epsilon_{10}^2)^{1/2}$
and $\bfepsilon_2= ({\bf e}_+ + \epsilon_{2-}{\bf e}_- + \epsilon_{20} {\bf e}_0)/(1+\epsilon_{2-}^2 + \epsilon_{20}^2)^{1/2}.$ 
Referring to Fig. \ref{fig.rabi2upper} and limiting ourselves to transition amplitudes that are linear in the small parameters $\epsilon$ we must account for the following undesired transitions starting from $|a\rangle=|10\rangle$: 
\begin{eqnarray}
|a\rangle\rightarrow |1-1\rangle &&\epsilon_{10}({\mathcal E}_1 {\bf e}_0 )({\mathcal E}_1 {\bf e}_+)^*
+\epsilon_{20}({\mathcal E}_2 {\bf e}_0 )({\mathcal E}_2 {\bf e}_+)^*\nonumber\\
|a\rangle\rightarrow |11\rangle &&\epsilon_{10}^*({\mathcal E}_1 {\bf e}_+ )({\mathcal E}_1 {\bf e}_0)^*+\epsilon_{20}^*({\mathcal E}_2 {\bf e}_+ )({\mathcal E}_2 {\bf e}_0)^*\nonumber\\
|a\rangle\rightarrow |2-2\rangle &&\epsilon_{1-}({\mathcal E}_1 {\bf e}_- )({\mathcal E}_2 {\bf e}_+)^* 
\nonumber\\ 
|a\rangle\rightarrow |2-1\rangle &&\epsilon_{10}({\mathcal E}_1 {\bf e}_0 )({\mathcal E}_2 {\bf e}_+)^* \nonumber\\  
|a\rangle\rightarrow |21\rangle &&\epsilon_{20}^*({\mathcal E}_1 {\bf e}_+ )({\mathcal E}_2 {\bf e}_0)^* 
\nonumber\\ 
|a\rangle\rightarrow |22\rangle &&\epsilon_{2-}^*({\mathcal E}_1 {\bf e}_+ )({\mathcal E}_2 {\bf e}_-)^* 
\label{eqs.leaka}
\end{eqnarray}
and the following transitions starting from $|b\rangle=|20\rangle$: 
%
\begin{eqnarray}
|b\rangle\rightarrow |1-1\rangle &&\epsilon_{20}({\mathcal E}_2 {\bf e}_0 )({\mathcal E}_1 {\bf e}_+)^*\nonumber\\
|b\rangle\rightarrow |11\rangle &&\epsilon_{10}^*({\mathcal E}_2 {\bf e}_+ )({\mathcal E}_1 {\bf e}_0)^*\nonumber\\
|b\rangle\rightarrow |2-2\rangle &&\epsilon_{1-}({\mathcal E}_1 {\bf e}_- )({\mathcal E}_1 {\bf e}_+)^*
                                   +\epsilon_{2-}({\mathcal E}_2 {\bf e}_- )({\mathcal E}_2 {\bf e}_+)^*\nonumber\\
|b\rangle\rightarrow |2-1\rangle &&\epsilon_{10}({\mathcal E}_1 {\bf e}_0 )({\mathcal E}_1 {\bf e}_+)^*
                                   +\epsilon_{20}({\mathcal E}_2 {\bf e}_0 )({\mathcal E}_2 {\bf e}_+)^*\nonumber\\
|b\rangle\rightarrow |21\rangle &&\epsilon_{10}^*({\mathcal E}_1 {\bf e}_+ )({\mathcal E}_1 {\bf e}_0)^*
                                   +\epsilon_{20}^*({\mathcal E}_2 {\bf e}_+ )({\mathcal E}_2 {\bf e}_0)^*\nonumber\\
|b\rangle\rightarrow |22\rangle &&\epsilon_{1-}^*({\mathcal E}_1 {\bf e}_+ )({\mathcal E}_1 {\bf e}_-)^*
                                   +\epsilon_{2-}^*({\mathcal E}_2 {\bf e}_+ )({\mathcal E}_2 {\bf e}_-)^*\nonumber\\&&
\label{eqs.leakb}
\end{eqnarray}
We have labelled the kets as $|F,m_F\rangle$ and the factors on the right indicate the participating fields 
and polarizations. The Clebsch-Gordan factors and detunings associated with these transitions are different than those that 
apply for the desired $|a\rangle\rightarrow|b\rangle$ transition. We will assume that the  fields are in two-photon 
resonance for the $|a\rangle\rightarrow|b\rangle$ transition accounting for the ac Stark shifts given in Table \ref{tab.acramanstark}.
The Rabi frequencies normalized to 
that of the desired transition are given in Table \ref{tab.polarization} which  shows the transition amplitude  for excitation of an undesired state 
with parameters that give a $\pi$ pulse
on the desired transition.

\begin{widetext}

\begin{table}[!t]
\centering
\begin{tabular}{|l||c|c|r|}
\toprule
           &                           & transition Stark & transition \\
transition & $K$ & shift [MHz]& amplitude \\
\colrule
    $|a\rangle\rightarrow |1-1\rangle$& $-\frac{2\epsilon_{10}}{ (\Delta_{11}-\Delta_e)}-\frac{2\epsilon_{20}}{ (\Delta_{11}-\Delta_e-\omega_{ba})}$ & -2.41&$-4.1\times 10^{-3}$\\
   $|a\rangle\rightarrow |11\rangle$ & 
$\epsilon_{10}^*\left(\frac{1}{\Delta_{11}} - \frac{3}{\Delta_{11}-\Delta_e}\right)
+\epsilon_{20}^*\left(\frac{1}{\Delta_{11}-\omega_{ba}} - \frac{3}{\Delta_{11}-\Delta_e-\omega_{ba}}\right)$ & 2.42&$-4.1\times 10^{-3}$\\
     $|a\rangle\rightarrow |2-2\rangle$&  $-\sqrt6 \epsilon_{1-}\left(\frac{1}{\Delta_{11}} - \frac{1}{\Delta_{11}-\Delta_e}\right)$ & -4.79&$1.3\times 10^{-5}$\\
     $|a\rangle\rightarrow |2-1\rangle$& $\frac{2\sqrt3 \epsilon_{10}}{\Delta_{11}-\Delta_e}$ & -2.39&$3.4\times 10^{-3}$\\
     $|a\rangle\rightarrow |21\rangle$&  $-\sqrt3 \epsilon_{20}^*\left(\frac{1}{\Delta_{11}} + \frac{1}{\Delta_{11}-\Delta_e}\right)$  & 2.38&$-3.4\times 10^{-3}$\\
     $|a\rangle\rightarrow |22\rangle$& $\sqrt6\epsilon_{2-}^*\left(\frac{1}{\Delta_{11}} - \frac{1}{\Delta_{11}-\Delta_e}\right)$  & 4.75&$-1.4\times 10^{-5}$\\ 
$|b\rangle\rightarrow |1-1\rangle $ &$-\frac{2\epsilon_{20}}{ \Delta_{11}}$  &-2.41 &$-1.9\times 10^{-3}$\\
 $|b\rangle\rightarrow |11\rangle $ &  $-\epsilon_{10}^*\left(\frac{1}{\Delta_{11}} - \frac{3}{\Delta_{11}-\Delta_e}\right)$  &
 2.42&$2.0\times 10^{-3}$\\
  $|b\rangle\rightarrow |2-2\rangle$&
$-\sqrt6\epsilon_{1-}\left(\frac{1}{\Delta_{11}+\omega_{ba}} - \frac{1}{\Delta_{11}-\Delta_e+\omega_{ba}}\right)
-\sqrt6\epsilon_{2-}\left(\frac{1}{\Delta_{11}} - \frac{1}{\Delta_{11}-\Delta_e}\right)
$  & -4.79&$2.5\times 10^{-5}$\\
 $|b\rangle\rightarrow |2-1\rangle $ & $\frac{2\sqrt3\epsilon_{10}}{\Delta_{11}+\omega_{ba}} +\frac{2\sqrt3\epsilon_{20}}{\Delta_{11}} $&-2.39 &$6.5\times 10^{-3}$\\ 
     $|b\rangle\rightarrow |21\rangle$&
$\sqrt3\epsilon_{10}^*\left(\frac{1}{\Delta_{11}+\omega_{ba}} + \frac{1}{\Delta_{11}-\Delta_e+\omega_{ba}}\right)
+\sqrt3\epsilon_{20}^*\left(\frac{1}{\Delta_{11}} + \frac{1}{\Delta_{11}-\Delta_e}\right)
$   &2.38 &$6.6\times 10^{-3}$\\
$|b\rangle\rightarrow |22\rangle $ &
$-\sqrt6\epsilon_{1-}^*\left(\frac{1}{\Delta_{11}+\omega_{ba}} - \frac{1}{\Delta_{11}-\Delta_e+\omega_{ba}}\right)
-\sqrt6\epsilon_{2-}^*\left(\frac{1}{\Delta_{11}} - \frac{1}{\Delta_{11}-\Delta_e}\right)
$   & 4.75&$2.5\times 10^{-5}$\\
\botrule
\end{tabular}
\caption{Transition amplitudes out of the computational basis due to imperfect optical polarization. 
The effective Raman frequencies  are given by $\Omega_R= ( 2 e^2 R_{5S,5P_{1/2}}^2/(\epsilon_0 c\hbar^2))(K/72)I$, with $I$ the intensity of each Raman beam. The last column shows the amplitude of the leakage state for a $\pi$ pulse on the $|a\rangle\rightarrow|b\rangle$ (or $|b\rangle\rightarrow |a\rangle$) transition, evaluated using Eq. (\ref{eq.pipulse}) with $I=100~\mu\rm W$ and $\Delta_{11}/2\pi=-100~\rm GHz$ such that $|\Omega_R(|a\rangle\rightarrow|b\rangle)|/2\pi=4.6~\rm MHz.$
}
\label{tab.polarization}
\end{table}
\end{widetext}

\end{document}